\documentclass[%
 reprint,
 amsmath,amssymb,
 aps,
 pra,
]{revtex4-1}

\usepackage{graphicx}
\usepackage{dcolumn}
\usepackage{bm}
\usepackage[overload]{empheq}
\usepackage{cases} 

\begin{document}

\preprint{APS/123-QED}

\title{Nash Equilibria in the Response Strategy of Correlated Games}

\author{A.D. Correia}
 \email{Corresponding author: a.duartecorreia@uu.nl}
\author{H.T.C. Stoof}%

\affiliation{%
Institute for Theoretical Physics and Center for Complex Systems Studies, Utrecht University, P.O. Box 80.089, 3508 TB, Utrecht, The Netherlands.
}%

\date{\today}

\begin{abstract}
In nature and society problems arise when different interests are difficult to reconcile, which are modeled in game theory. While most applications assume uncorrelated games, a more detailed modeling is necessary to consider the correlations that influence the decisions of the players. The current theory for correlated games, however, enforces the players to obey the instructions from a third party or "correlation device" to reach equilibrium, but this cannot be achieved for all initial correlations. We extend here the existing framework of correlated games and find that there are other interesting and previously unknown Nash equilibria that make use of correlations to obtain the best payoff. This is achieved by allowing the players the freedom to follow or not to follow the suggestions of the correlation device. By assigning independent probabilities to follow every possible suggestion, the players engage in a response game that turns out to have a rich structure of Nash equilibria that goes beyond the correlated equilibrium and mixed-strategy solutions. We determine the Nash equilibria for all possible correlated Snowdrift games, which we find to be describable by Ising Models in thermal equilibrium. We believe that our approach paves the way to a study of correlations in games that uncovers the existence of interesting underlying interaction mechanisms, without compromising the independence of the players. 
\end{abstract}

\maketitle




\textit{Introduction - }Game theory \cite{fudenberg1991game} has been used as a powerful tool to model problems from diverse research areas, such as biology \cite{smith1973logic,smith1982evolution,nowak2004evolutionary}, economics \cite{kreps1990game,van2016non}, politics \cite{morrow1994game} and social sciences \cite{buskens2016effects}. Many applications involve uncorrelated coordination games \cite{maclean2006resource,rand2011dynamic}, such as the Prisoner’s Dilemma \cite{gracia2012heterogeneous,turner1999prisoner} and the Snowdrift games \cite{gore2009snowdrift, zomorrodi2017genome}, where the players make their decisions independently. The Snowdrift game is of particular interest because has it been shown to model biological conflict \cite{gore2009snowdrift}, yet the optimal solution of this game is reached when the players are allowed to communicate, using correlations. Thus, to further advance the applications of games to real life situations, a more complex treatment is required \cite{oliveira2014evolutionary,pollock1994social}. In that direction, an improvement can be made if we do not to assume from the start that a game is uncorrelated, but consider that the decisions that the players make can be informed both by the payoff function and by underlying correlations in the system. A prominent example is that of Evolutionary Stable Strategies \cite{smith1973logic}: while the analysis has been successful at describing actions at the level of what the outcomes are, namely the phenotype, what originates these behaviors, namely the connections with the genotype, is a field of active research \cite{nowak2004evolutionary} and may lie precisely in introducing correlations. Another area that has seen much interest recently consists of having several players playing games on a network against each other, and so they must make a choice to optimize their payoff. If the players are correlated we expect that we can describe this by an Ising model in a magnetic field, with the correlated probabilities corresponding to a Boltzman distribution. For all these diverse applications, it is thus crucial that the correlations do not completely determine the actions, but merely inform them. To this end, we analyze what outcomes arise when we give the players the freedom to act on the correlations. This is an improvement on the current theory of correlated games, which has a much more limited action range. Furthermore, we describe how to introduce this freedom for each player independently using the Ising model.
\\

\begin{figure}[!ht]
\centering
\includegraphics[width=.8\linewidth]{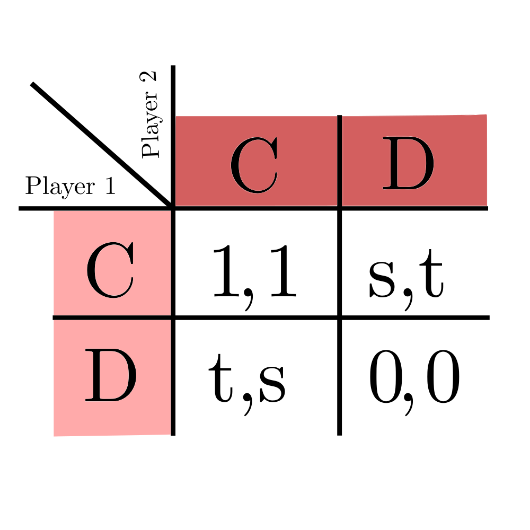}
\caption{Normalized payoff table for two-by-two, symmetric coordination games.}
\label{fig:table}
\end{figure}

We consider, without loss of generality, symmetric, two by two, coordination games. The players have access to the payoff matrix in Fig. \ref{fig:table}, that settles how much reward each player receives given the actions of all the players. Player \( 1 \) receives the payoffs on the left side of the comma, while player two receives the payoff on the right side of the comma, dependent on the two strategies chosen. The different games are defined by the range of the parameters: the Harmony game has \( 0<s<1 \) and \( 0<t<1 \); the Stag-Hunt game has \( -1<s<0 \) and \( 0<t<1 \); the Prisoner’s Dilemma has \( -1<s<0 \) and \( 1<t<2 \); and the Snowdrift game, also known as Chicken or Hawk-Dove game, has \( 0<s<1 \) and \( 1<t<2 \). Depending on the game being played, the players decide on the best strategy based on how much they will win given all the possible strategies of the adversaries. For the uncorrelated case, the objective is to maximize the expected payoff by assigning a probability \( P_{C} \) to playing \( C \) (to cooperate), so that \( D \) (to defect) is played with the probability \( 1-P_{C} \). A Nash equilibrium \cite{nash1950equilibrium} is reached for a strategy, i.e., a value of \( P_{C} \), that none of the players wants to deviate from. If \( P_{C} \) is \( 0 \) or \( 1 \), it is a pure Nash equilibrium, otherwise it is a mixed strategy equilibrium. The mixed strategy solution is of particular importance in the Snowdrift games. This game has two pure Nash equilibria in which the players adopt opposite pure strategies, but these cannot be reached without introducing correlations between the players. Therefore, the best solution is a mixed strategy Nash equilibrium, where the probability of playing \( C \) for each player is \( P_{C}^{\ast}=s/ \left( t+s-1 \right)  \).

This solution for the Snowdrift game is the best that the players can do without communicating, but, for most parameters, better results are obtained if they can both play opposite strategies systematically. This, however, requires the introduction of correlations between the players. To illustrate this, we consider the extreme example of a simple traffic light on a cross-road. The cross-road can also be described as a Snowdrift game, where the best payoffs are obtained when one of the drivers decides to stop (to cooperate) and the other decides to go (to defect). Since the players cannot communicate, a traffic light is needed to achieve the optimal situation. The traffic light can be seen as the correlation device, which assigns a publicly known probability \( p_{ \mu  \nu } \) to a certain state \(  \mu  \nu  \), with Greek indices taking the values \textit{C} and \textit{D}. In this particular example, the device assigns equal probabilities to the states \( CD \) and \( DC \), while assigning zero probability to \( CC \) and \( DD \). Since the correlation is very strong and there is a big penalty if both players defect simultaneously, the players always want to follow the correlation device, and the game is thus in a "correlated equilibrium". A general correlation device, however, assigns non-zero probabilities to all the possible outcomes of a game. If this is the case for the states with low payoff, the question arises whether always following is the best strategy for the players. According to the existing theory \cite{aumann1987correlated}, they should always follow the correlation device if they are in correlated equilibrium, and they should fall back to the uncorrelated mixed-strategy solution otherwise. This ensures that the probabilities in correlated equilibrium coincide with the final distribution of outcomes, such that they represent the actual statistics of the game. Another way of representing any symmetric, and thus possibly correlated, probability distribution over the four possible states is through a Boltzman distribution of an Ising system. Let each player be represented by a particle with spin, and let $C$ and $D$ correspond to the spin states "up" and "down", such that $\mu, \nu \in \{\uparrow, \downarrow \}$. We can rewrite the probabilities as

\begin{equation} \label{corrprobs}
p_{\mu \nu}= \frac{e^{-\beta H_{\mu \nu} }}{Z}, 
\end{equation} with $\beta$ the inverse of a generalized thermal energy, $H_{\mu \nu}$ the Hamiltonian of the Ising system, decomposable into the sum of a constant term that can be absorbed into the partition function, an Ising term and a Zeeman term \cite{stoof2009ultracold}, and $Z=\sum_{\mu,\nu} e^{-\beta H_{\mu\nu}}$ the partition function. Thus, if the players always follow the instructions from the correlation device, this distribution perfectly describes the actions of the players.

\begin{figure*}[!ht]
\centering
\includegraphics[width=15cm,height=7.9cm]{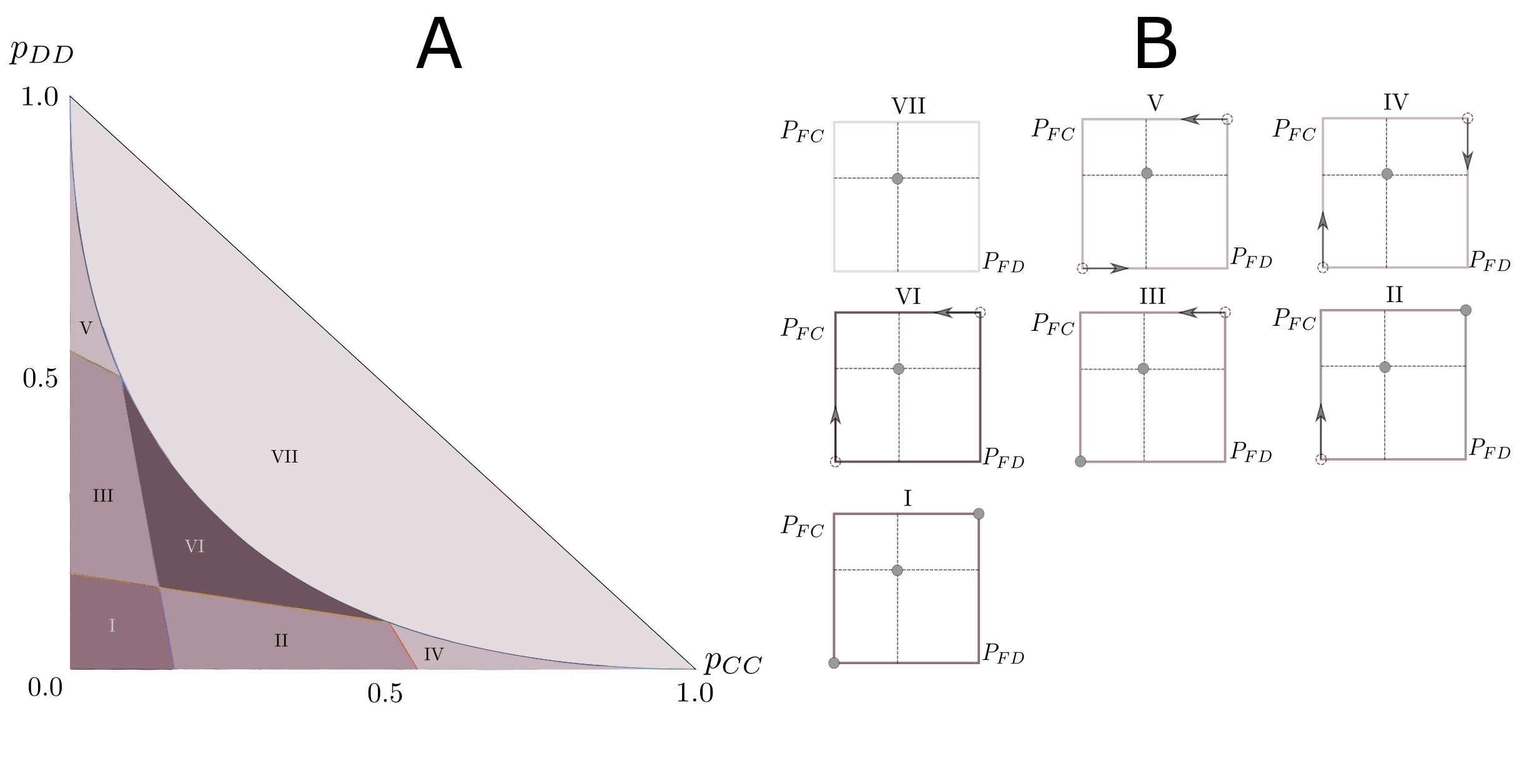}
\caption{Symmetric correlations and associated equilibrium response strategies for the Snowdrift game with $s = 0.5$ and $t = 1.2$.
\textbf{A}, Illustration of all correlated Nash equilibria of the Snowdrift game in the \( P_{CC}-P_{DD} \) plane, for parameters representative of \( s>t-1 \). \textbf{B}, Schematic representation of the equilibrium value of the response probabilities in the \( P_{FD}- P_{FC} \) plane that can be found in each region enumerated in \textbf{A}. The lines that separate each region in \textbf{A} are obtained by imposing a particular sign on a slope of a response probability and using the associated value for that probability. Between each straight line and the curved line, given by \( p_{DD}= 1+p_{CC}+2\sqrt[]{p_{CC}} \), there exists a solution with one of the slopes strictly greater or smaller than zero and the other one equal to zero, such that one of the response probabilities is \( 0 \) or \( 1 \) and we can find a value for the other probability that lies between these values. Each response probability associated with a zero slope can have values that range between the value of the other probability in the extreme and its associated mixed-strategy solution (due to the limiting condition that \( p_{DD} \leq 1-p_{CC} \)), represented in \textbf{B} by the dotted line. Bellow the lines, both slopes have the same sign and both probabilities are in the same extreme of the interval. The arrows in \textbf{B} depict how the value of the probabilities change as we move from the straight lines towards the curved line, which when reached sends the probabilities to the uncorrelated mixed-strategy solution that is always a solution in all regions. The upper line that delimitates regions I and II and the rightmost line that delimitates region II are the lines that arise when both slopes are positive. They correspond to the correlated equilibrium conditions. The intersection of these two lines is the mixed-strategy solution when written as a response strategy. Thus in these two regions the correlated equilibrium is a solution, although it is not unique or always optimal in payoff. Moreover, there are regions where the correlated equilibrium is unstable but other solutions exist that make use of the correlations to increase the payoff. For \( s<t-1 \) the results are similar, but there occurs a swapping among the lines: the rightmost boundary of region I becomes the leftmost, and the top boundary of region III becomes the bottom one, which changes the equilibria in regions II, III and VI correspondingly. When \( s=t-1 \) these boundaries overlap and these regions disappear.}\label{fig:equilibria}
\end{figure*}

\begin{figure*}[!ht]
\centering
\includegraphics[width=15cm,height=6.1cm]{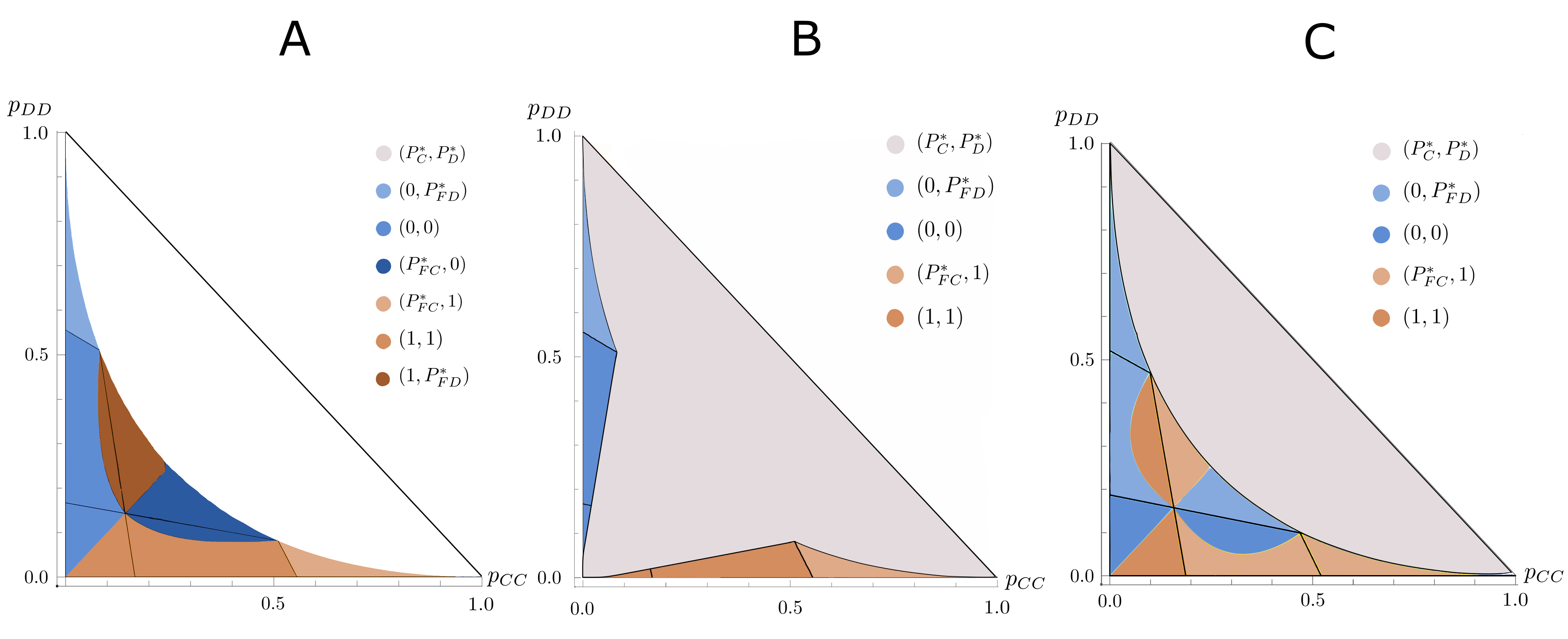}
\caption{Equilibrium response strategies with highest payoff per region, for different parameters. \textbf{A}, Equilibria corresponding to highest payoff by region, for \( s=0.5 \) and \( t=1.2 \) \( \left( s>t-1 \right)  \), without the mixed-strategy solution. \textbf{B}, Equilibria corresponding to highest payoff, for the same parameters as \textbf{A}, including the mixed-strategy solution. \textbf{C}, Equilibria corresponding to highest payoff by region, for \( s=0.23 \) and \( t=1.5  \left( s<t-1 \right) \), including the mixed-strategy solution. The existing equilibrium solutions are compared within a region, according to the description in fig. \( 2 \). The darkening of the colors represents a higher absolute value of the payoff when compared to the best payoffs of the neighboring regions, but the actual value changes within the region, except for the mixed-strategy solution, of which the is constant. All the payoffs corresponding to the best solution are connected to one another. As the parameters vary, the blue solution in region \( II \) and the orange solution in region \( III \) either increases or decreases in area. In \textbf{b} we see that the mixed strategy can be the best solution for a game with strong off-diagonal correlations, but this is not generally the case for all parameters. In the regions where the correlated equilibrium exists, there is, for certain initial probabilities, a better alternative solution. All the renormalized games will correspond to final probabilities where the correlated equilibrium is the best solution, such that the players do not want to deviate anymore from their chosen strategy. Because the mixed-strategy\ payoff is a constant, we see that correlations always provide a payoff that is at least as good.}\label{fig:payoff}
\end{figure*}

To show that this is not the complete picture, we now allow the players to deviate from the instructions of the correlation device in a controlled manner. The decisions to follow or not to follow the instructions become the new actions that the players can take, while they are still not able to coordinate. To implement this, each player \( i \) can follow with probability \( P_{F \mu }^{i} \), and thus not follow with probability \(  P_{NF \mu }^{i}=1-P_{F \mu }^{i} \) the instruction \(   \mu  \) that they receive. These we call the "response probabilities". The renormalized probability \( p_{ \mu  \nu }^{R} \) that a certain final state \(  \mu  \nu  \) is reached is given by the sum over the initial probability distribution weighted by the probability that the initial states \(  \mu' \nu' \) gets converted to a specific final state \(  \mu  \nu  \) through the players' response. Hence \cite{mailath1997correlated,wong2004evolutionarily,lenzo2006correlated,metzger2018evolution}

\begin{equation} \label{renormprobs}
p_{ \mu  \nu }^{R}= \sum _{ \mu',  \nu'}^{}P_{ \mu  \leftarrow  \mu '}^{1}P_{ \nu  \leftarrow   \nu'}^{2} p_{ \mu' \nu'} ,
\end{equation} with  \( P_{ \mu  \leftarrow  \mu'}^{i} \) the probability that player \( i  \) is told to play \(   \mu '  \) but plays \(  \mu  \). As an example, the probability that the final state is \( CC \) is now\par
\begin{align}
&p_{CC}^{R}= P_{FC}^{1}P_{FC}^{2}p_{CC} + P_{FC}^{1}P_{NFD}^{2}p_{CD}  \nonumber \\ 
& + P_{NFD}^{1}P_{FC}^{2}p_{DC}+P_{NFD}^{1}P_{NFD}^{2}p_{DD}.
\end{align} 
The expected payoff of a player is given by the payoffs averaged over the renormalized probabilities, which depends linearly on the response probabilities of that player as\par
\begin{equation}\label{payoffequation}
	 \langle u^{i} \rangle = \sum _{ \mu , \nu }^{}u_{ \mu  \nu }^{i} p_{ \mu  \nu }^{R}=C_{C} P_{FC}^{i}+C_{D} P_{FD}^{i}+C_{E},
\end{equation} with  \( u_{ \mu  \nu }^{i} \)  the payoff of player\textit{  \( i \) } in the state  \(  \mu  \nu  \). Here the coefficients \( C_{C} \), \( C_{D} \) and \( C_{E} \) depend linearly on the initial correlation probabilities and on the response probabilities of the other player.\par

A Nash equilibrium in the response strategy is achieved if there is no incentive for player \( i \) to change the probabilities \( P_{F \mu }^{i} \). This is achieved by imposing that the slope of the expected payoff with respect to each response probability, either \( C_{C} \) or \( C_{D} \) in the evaluation of the total player's payoff, is zero, unless the equilibrium response probability is 0 or 1, in which case the slope should be negative or positive, respectively. The intuition is that equilibrium is reached when the payoff of the players cannot be improved anymore by changing their own response probabilities while keeping those of the other players fixed at the equilibrium values. The independence of this analysis separately for each response probability represents Bayes rationality \cite{aumann1987correlated} of the players towards the final states given the initial information that they receive. \par

As a result, there are three possible Nash equilibria for each response probability: \( P_{F \mu }=0 \), \( P_{F \mu }=1 \) and \( 0<P_{F \mu }<1 \).\  In our two by two coordination games, this amounts to nine possible types of equilibrium response strategies, but which ones are actually realized depends on the payoff parameters. We find that the conditions where "always follow", i.e., \( P_{F \mu }=1 \) is a stable solution, correspond to the Bayes rational conditions of the correlated equilibrium, indicating that this is only one possible response equilibrium. However, each renormalized set of probabilities generates a new correlated game for which the response equilibrium exactly matches a correlated equilibrium, from which the players by definition indeed do not want to deviate. Using the slopes \( C_{C} \) and \( C_{D} \) to evaluate the Nash equilibria, each of the response probabilities has to be subject to one of the three above-mentioned conditions simultaneously. To guarantee that none of the players wants to deviate, each slope is calculated assuming that the response probabilities of the other players are in equilibrium, such that a self-consistent solution is obtained.\par

We studied the response strategies of the Snowdrift game for all initial (symmetric) correlation probabilities and show the results in Fig. \ref{fig:equilibria}. In addition to the expected existence of a correlated equilibrium region, where interestingly also other Nash equilibria exist, we see other regions with other types of equilibria that use correlations to optimize the payoff. The mixed strategy of the uncorrelated Snowdrift game is always an equilibrium solution in every region. 

To choose the best response for each region, we compare in Fig. \ref{fig:payoff} the payoffs of all the possible response equilibria. In the correlated equilibrium region, we see that there sometimes exist other solutions that have better payoffs. This can be intuitively understood from our cross-road example where the strategy "always not follow"  has identical payoff to "always follow".

Having found the optimal response probabilities, we can now incorporate them in the Ising model that effectively describes the final statistics of the game. Using the language of statistical physics, the response probabilities in equilibrium can be written as 

\begin{equation} \label{response}
P^i_{\mu \leftarrow \mu'}= \frac{e^{-\beta B^i_{\mu \leftarrow \mu'} }}{Z^i_{\mu'}},
\end{equation} with the partition function given by

\begin{equation}\label{partfunc}
Z^i_{\mu'}=\sum_{\mu} e^{-\beta B^i_{\mu \leftarrow \mu'}},
\end{equation} where $B^i_{\mu \leftarrow \mu'}$ are the appropriate energies that are explicitly given in the Methods section. The renormalized correlated probabilities, by eq.\ref{corrprobs}, are written in closed form as

\begin{equation} \label{corrprobsR}
p^R_{\mu \nu}= \frac{e^{-\beta H^R_{\mu \nu} }}{Z^R}.
\end{equation} Using eqs.\ref{corrprobs}, \ref{response} and \ref{corrprobsR}, we are able to describe the renormalized Ising Hamiltonian as

\begin{equation} \label{newising}
H^R_{\mu \nu}= -\frac{1}{\beta} \ln \left(\sum_{\mu' \nu'} Z^1_{-\mu'} Z^2_{-\nu'}  e^{-\beta \left(B^1_{\mu \leftarrow \mu'} + B^2_{\nu \leftarrow \nu'} + H_{\mu' \nu'} \right)} \right).
\end{equation} The minus sign indicates the opposite play, i.e., $-C=D$ and $-D=C$, or the corresponding interchange between spin states "up" and "down". The renormalized Hamiltonian is obtained by the players choosing the energy parameters $B^i_{\mu \leftarrow \mu'}$ independently, which in this formalism is what allows the players to have complete control over the final probabilities. How eqs.\ref{response} and \ref{newising} are obtained is explicitly explained in the Methods section. Note that if eq.\ref{newising} is interpreted as a renormalization-group transformation, the Nash equilibria correspond to the fixed points of this transformation \cite{stoof2009ultracold}.  
\\

\textit{Discussion - } The introduction of the response strategy to correlated games opens up several new features. We showed that the correlated equilibrium is only a particular response equilibrium, but that other Nash equilibria exist. These new equilibria renormalize to a correlated equilibrium even if the initial game is out of correlated equilibrium, showing that the players even then can still use the correlations to achieve a better payoff. The extra information in the correlations is two-fold: either the final distributions of outcomes informs us about an underlying correlation structure, or the players can independently improve on externally imposed initial correlations, motivated by stability and payoff maximization. Regarding evolutionary game theory, the correlated Evolutionary Stable Strategy \cite{wong2004evolutionarily,cripps1991correlated} is, similarly, only one equilibrium where the agents always follow the correlations, which suggests that the evolutionary stable solution is not unique. Other possible applications involve modeling emergent behavior when games are played on networks \cite{buskens2016effects,rand2011dynamic,gracia2012heterogeneous,cassar2007coordination,broere2017network}. While all the related research relies on numerical methods, our approach may provide some analytical insight to the results, since for the two-player game the renormalized probabilities of the optimal Nash equilibrium are equivalent to an Ising Model in a magnetic field. We show how the players can introduce an energy parameter to change the Hamiltonian to effectively obtain the renormalized probabilities.  This suggests that statistical-physics tools can be used to model a simple network \cite{adami2018thermodynamics}, but it remains as an interesting open question how well this would describe the non-local network effects. Bridging the gap between correlated and uncorrelated games will also prove useful to better model decision-making in economics, since the response probabilities allow us to include interactions between agents that influence their decisions.

\onecolumngrid

\appendix

\section{Slope analysis for the Snowdrift Game}

In eq.\ref{payoffequation} we have an expression for the expected payoff of player 1, with the coefficients \( C_{C} \) and \( C_{D} \). To make the various dependencies clearer, we now denote these explicitly as \( C_{C} \left( p_{CC}, p_{DD}, P_{FC}^{2}, P_{FD}^{2} \right)  \) and \( C_{D} \left( p_{CC}, p_{DD}, P_{FC}^{2}, P_{FD}^{2} \right)  \), respectively. \par

For the Snowdrift game, which is a symmetric game, it is natural to assume also a symmetric probability distribution, so \( p_{CD}=p_{DC}= \left( 1-p_{CC}- p_{DD} \right) /2 \). When both slopes are positive, the equilibrium will be at \( P_{FC}^{1}=1 \) and \( P_{FD}^{1}=1 \), and due to the symmetry of the game, player \( 1 \) will not want to change these probabilities when player \( 2 \) has the same equilibrium probabilities. To reach equilibrium, the conditions thus become\par

\begin{equation}\label{methods1}
	 \left\{ \begin{matrix}
	C_{C} \left( p_{CC}, p_{DD},1,1 \right) >0,\\
	C_{D} \left( p_{CC}, p_{DD},1,1 \right) >0.
	\end{matrix}\right.
\end{equation}

These conditions are equivalent to the correlated equilibrium conditions. Similarly, a second kind of equilibrium is reached when the slopes are negative and the response probabilities are zero, i.e.,

\begin{equation}\label{methods2}
	 \left\{ \begin{matrix}
	C_{C} \left( p_{CC}, p_{DD},0,0 \right) <0,\\
	C_{D} \left( p_{CC}, p_{DD},0,0 \right) <0.
	\end{matrix} \right.
\end{equation}
A third type of equilibrium exists when both slopes equate to zero

\begin{equation}\label{methods3}
	  \left\{\begin{matrix}
	C_{C} \left( p_{CC}, p_{DD},P_{FC}^{1\ast} ,P_{FD}^{1\ast} \right) =0,\\
	C_{D} \left( p_{CC}, p_{DD},P_{FC}^{1\ast} ,P_{FD}^{1\ast} \right) =0,
	\end{matrix} \right.
\end{equation} for which the solution coincides with the mixed-strategy equilibrium solution, with \( P_{FC}^{1\ast}=P_{FC}^{2\ast}=P_{C}^{\ast} \) and \( P_{FD}^{1\ast}=P_{FD}^{2\ast}=P_{D}^{\ast} \).
The last type of equilibria comes in four possible guises, consisting of one of the conditions being zero, while the other is strictly positive or negative. For instance, we can have

\begin{subequations}
\begin{align}[left = \empheqlbrace\,]
      & C_{C} \left( p_{CC}, p_{DD},1 ,P_{FD}^{1\ast} \right) >0 \label{methods41}\\
       & C_{D} \left( p_{CC}, p_{DD},1 ,P_{FD}^{1\ast} \right) =0 \label{methods42}.
    \end{align}
\end{subequations} If we calculate the specific value of the response probability using eq.\ref{methods42}  and substitute this value in eq.\ref{methods41}, a new condition arises for symmetric games, namely

\begin{equation}
	p_{DD}< 1+p_{CC}-2\sqrt[]{p_{CC}}.
\end{equation} Under this condition, and \( p_{DD} \leq  1-p_{CC} \), each equilibrium response probability that has a zero slope in the expected payoff, has a limited range of possible values, which, depending on the sign of the associated inequality, goes from \( 1 \) or \( 0 \) to \( P_{C}^{\ast} \) or \( P_{D}^{\ast} \). Each of the four possible combinations correspond to one of the four response equilibria of this kind. The solution of the example given above is \( P_{FC}^{1\ast}=1  \) and \(  P_{D}^{\ast} \) \( <P_{FD}^{1\ast}< \) 1, with the specific value of \( P_{FD}^{1\ast} \) depending on the initial correlated probabilities. \par
Due to the value of the Snowdrift game's payoff parameters, the two other conditions that would arise from having the two conditions with opposite strict inequalities do not have solutions, so only seven equilibria exist in total for this game.\par
Note that in a game-theoretical notation, each of the above equilibrium conditions can be summarized as

\begin{equation*}
\sum _{ \mu , \nu , \nu '}^{}u_{ \mu  \nu }^{1} P_{ \mu  \leftarrow  \mu '}^{1\ast}P_{ \nu  \leftarrow  \nu '}^{2\ast}p_{ \mu ' \nu ' } \geq  \sum _{ \mu , \nu , \nu '}^{}u_{ \mu  \nu }^{1} P_{ \mu  \leftarrow  \mu '}^{1}P_{ \nu  \leftarrow  \nu '}^{2\ast}p_{ \mu ' \nu ' }.
\end{equation*} These comprehend two conditions, one for every value of \(  \mu ' \). Summing over these shows that the expected payoff of player \( 1 \) cannot be improved by deviating from the equilibrium strategy, which is the requirement for a Nash equilibrium. The stronger statement that both conditions are satisfied separately expresses the Bayes rationality of player \( 1 \).\

\section{Renormalized Ising Model}

To insert the response probabilities in the Ising model, each reaction from the players will have a Zeeman-like energy: either $-B_\mu^i$ if they follow $\mu$, or $+B_\mu^i$ if they do not follow:

\begin{equation} \label{renormmags}
B^i_{\mu \leftarrow \mu'}= -\delta_{\mu \mu'} B^i_\mu + (1-\delta_{\mu \mu'})B^i_\mu.
\end{equation} We can then rewrite the response probabilities as

\begin{align}
P^i_{\mu\leftarrow \mu'} &= \frac{ \delta_{\mu \mu'} e^{\beta B^i_{\mu'}} + (1-\delta_{\mu \mu'}) e^{-\beta B^i_{\mu'}}}{Z_{\mu'}^i} = \frac{e^{-\beta B^i_{\mu \leftarrow \mu'}}}{Z_{\mu'}^i}, \label{newtransprob} 
\end{align} with $Z_{\mu'}^i$ as given in eq.\ref{partfunc}. 
Using eq.\ref{renormprobs}, we calculate the renormalized Ising energies:

\begin{align}
&H^R_{\mu \nu} = -\frac{1}{\beta} \ln \left(Z^R \; p^R_{\mu \nu} \right) = -\frac{1}{\beta} \ln \left(\sum_{\mu',\nu'} P^1_{\mu \leftarrow \mu'}P^2_{\nu \leftarrow \nu'} p_{\mu' \nu'} \right) - \frac{1}{\beta} \ln \left(Z^R\right) \nonumber
\end{align} Introducing eq.\ref{newtransprob} and the initial correlation probabilities with energies $H_{\mu' \nu'}$, with associated partition function $Z$, we get a simplified version of the renormalized energies:

\begin{align}
&H^R_{\mu \nu}= -\frac{1}{\beta} \ln \left(\frac{Z^R}{Z Z_{C}^1 Z_{D}^1 Z_{C}^2 Z_{D}^2} \right)  -\frac{1}{\beta} \ln \left(Z_{D}^1 Z_{D}^2 e^{-\beta \left(B^1_{\mu \leftarrow C} + B^2_{\nu \leftarrow C} + H_{CC} \right)}  \right. \nonumber \\
&\left. + Z_{D}^1 Z_{C}^2e^{-\beta \left(B^1_{\mu \leftarrow C} + B^2_{\nu \leftarrow D} + H_{CD} \right)} + Z_{C}^1 Z_{D}^2 e^{-\beta \left(B^1_{\mu \leftarrow D} + B^2_{\nu \leftarrow C} + H_{DC} \right)} \right. \nonumber \\
&\left. + Z_{C}^1 Z_{C}^2e^{-\beta \left(B^1_{\mu \leftarrow D} + B^2_{\nu \leftarrow D} + H_{DD} \right)} \right).
\label{fancyenergy}
\end{align}The first term in eq.\ref{fancyenergy} drops out because we have that 

\begin{equation} \label{renormpartfunc}
Z^R=Z Z_{C}^1 Z_{D}^1 Z_{C}^2 Z_{D}^2.
\end{equation} Hence, we can rewrite $H^R_{\mu\nu}$ as in eq.\ref{newising}. The fact that the renormalized partition function is a product of the various partition functions expresses that the actions of the players enter the renormalized Hamiltonian in an independent manner.

\section{Background on Game Theory}

\subsection{Strategic-Form Games}

A strategic form game is defined by three elements: the finite set $\mathcal{I}$ of $I$ players, with $\mathcal{I}=\{1,2,...,I\}$; the pure-strategy space $S_i$ for each player $i \in \mathcal{I}$, representing the plays that each player has available; and the payoff functions $u_i(s_i,s_{-i})$, denoting the gain of player $i$ if he plays $s_i \in S_i$ and the other players, denoted by $-i$, play $s_{-i} \in S_{-i}$. Besides the pure strategies, the players can play a mixed strategy, in which player $i$ plays the pure strategy $s_i$ with probability $\sigma_i(s_i)$. A pure strategy is a particular case of a mixed strategy, that assigns probability $1$ to a certain element of the pure-strategy space.

The players do not have access to what their opponents will play, so the rational player has to consider all his possible moves. Taking this into account, the Nash equilibrium guarantees that each player chooses a strategy from which they do not want to deviate. A mixed strategy profile $\sigma_i^*$ is the Nash equilibrium if for all players $i$ we have that their average payoff obeys
\begin{equation} \label{nash}
\langle u_i \rangle(\sigma_i^*,\sigma_{-i}^*) \geq \langle u_i \rangle(\sigma_i,\sigma_{-i}^*),
\end{equation}
with $\sigma_i \in \Sigma_i$ any element of the set of all possible mixed strategy profiles. Since a set of probabilities is convex and compact, it is enough to guarantee that
\begin{equation} \label{weaker}
\langle u_i \rangle(\sigma_i^*,\sigma_{-i}^*) \geq \langle u_i \rangle(s_i,\sigma_{-i}^*).
\end{equation} 
for all $s_i \in S_i$.

If the inequality is strict, a pure-strategy Nash equilibrium ensues. In symmetric, two by two, two strategy games, i.e., $S_i=\{C,D\}$ where $C$ denotes to cooperate and $D$ to defect, Nash equilibria are easy to categorize. For the Harmony Game (HG) both players cooperate; in the Prisoner's Dilemma (PD) both defect, and the Stag-Hunt game (SH) has both these two equilibria. The Snowdrift Game (SG), also called Chicken or Hawk-Dove, has two pure strategy Nash equilibria, where one of the players cooperates and the other defects, but these are impossible to achieve, due to the symmetry of the game. The best strategy for this game is a mixed-strategy equilibrium, which assigns an equal probability to cooperate to each player.

In the games that we will analyze, there are only two actions and two players, so $\sigma_{i}=P_{C_i}$ with $i=1,2$, and we rewrite the equilibrium in eq.\ (\ref{nash}) as
\begin{equation}
\langle u_1 \rangle (P^*_{C_1}, P^*_{C_2})\geq \langle u_1 \rangle(P_{C_1}, P^*_{C_2}).
\end{equation}
Expanding, we get in first instance

\begin{align*}
&P^*_{C_1} P^*_{C_2} u_1 (C,C) + P^*_{C_1} (1-P^*_{C_2}) u_1 (C,D) + (1- P^*_{C_1}) P^*_{C_2} u_1 (D,C) + (1-P^*_{C_1})(1- P^*_{C_2}) u_1 (C,C)\\
 \geq & P_{C_1} P^*_{C_2} u_1 (C,C) + P_{C_1} (1-P^*_{C_2}) u_1 (C,D) + (1- P_{C_1}) P^*_{C_2} u_1 (D,C) + (1-P_{C_1})(1- P^*_{C_2}) u_1 (C,C).
\end{align*}
Subtracting the left-hand side from the right-hand side gives

\begin{align}
&\left( P^*_{C_1} - P_{C_1} \right) \left( P^*_{C_2} u_1 (C,C) + (1-P^*_{C_2}) u_1 (C,D) \right) + \left( P_{C_1} - P^*_{C_1} \right)  \left( P^*_{C_2} u_1 (D,C) + (1-P^*_{C_2}) u_1 (D,D) \right)\geq 0,
\end{align}

which after rearrangement of the terms gives the desired result

\begin{equation}\label{generic}
\left( P^*_{C_1} - P_{C_1} \right) \left[ P^*_{C_2} \left(u_1 (C,C)-u_1 (D,C) \right) + (1-P^*_{C_2}) \left(u_1 (C,D) -u_1 (D,D) \right) \right] \geq 0.
\end{equation}

For this condition to hold, the coefficient of $\left( P^*_{C_1} - P_{C_1} \right)$ has to have the same sign as $\left( P^*_{C_1} - P_{C_1} \right)$ itself. Hence, if $P_{C_1}$ is bigger than $P^*_{C_1}$, the coefficient has to be non-negative, and otherwise, non-positive. The condition in eq.\ (\ref{generic}) has to hold for all values of $P_{C_1}$ and so, for the case when $P^*_{C_1}$ is not in one of the extremes of the interval, $\left( P^*_{C_1} - P_{C_1} \right)$ can be positive, negative or zero. The only way to ensure that the condition is always true is if it is always zero, meaning that the coefficient has to be zero or

\begin{align}\label{slope}
&P^*_{C_2} u_1 (C,C) + (1-P^*_{C_2})u_1 (C,D) = P^*_{C_2} u_1 (D,C) + (1-P^*_{C_2})u_1 (D,D).
\end{align}

We can thus calculate that the mixed strategy equilibrium probability $P^*_{C2}$ by solving
\begin{equation} \label{equilibrium}
\left\langle u_1 \right\rangle (1,P^*_{C_2})= \left\langle u_1 \right\rangle (0,P^*_{C_2}).
\end{equation}
If $P^*_{C_1}$ is in one of the extremes, then the condition becomes a strict inequality and we obtain a pure strategy equilibrium.


The conditions of the probabilities are those of Kakutani's Theorem, used by Nash to prove the existence of the fixed points that we now know as "Nash Equilibria" \cite{nash1950equilibrium}. Analyzing the slope of the probability $P_{C_1}$ in the same way as done in our paper for the response probabilities proves the same result.

\subsection{Correlated Games and Correlated Equilibrium}

Suppose that the players made some agreement beforehand about what they will play, or that there is some external information that both share even if they do not communicate. This idea is formalized by extending the game with a correlation device. This device draws one of the possible final states, the \textit{true state} $\omega \in \Omega$, with probability $p(\omega)$ and subsequently informs each player of what they should play to achieve the true state $\omega$. Player $i$ then has information $h_i(\omega) \in H_i$, that is, he knows what true states are possible given the information he received. The probability that each of these states is the true state is given by $p(\omega|h_i)$. In the case of the coordination game described above $\Omega=\{CC,CD,DC,DD\}$ and, for example, $h_1(CD)=C$ and $\{\omega|h_1\}=\{CC,CD\}$. This means that if the true state is $CD$, player $1$ is told to play $C$, at which point he knows that either $CC$ or $CD$ are the possible true states. A correlated strategy $s(\omega)$ represents what pure or mixed strategy each player adopts in the true state $\omega$. The probability distribution $s(\omega)$ is the same as the initial distribution $p(\omega)$ of plays if the players always play according to what they are told.

The players have Bayesian rationality, which means that if we look at the final action of a certain player, their payoff conditioned on the fact that that was the information that they received from the correlation device must be greater than the payoff they could have achieved if he played something else, given the same information:

\begin{equation}\label{conditioned correlation}
\sum_{\{\omega|h_i\}} u_i(s_i(\omega), s_{-i}(\omega)) p(\omega|h_i) \geq \sum_{\{\omega|h_i\}} u_i(\bar{s}_i(\omega), s_{-i}(\omega)) p(\omega|h_i),
\end{equation}

which must be valid for all players $i$, all information structures $h_i(\omega)$ and all pure strategies $\bar{s}_i(\omega)$. We say that the players are Bayes rational towards the state of the world $\omega$.

If we sum over $h_i(\omega)$ after having multiplied with the corresponding probability, we arrive at
\begin{equation}\label{correlation}
\sum_{\omega} u_i(s_i(\omega), s_{-i}(\omega)) p(\omega) \geq \sum_{\omega} u_i(\bar{s}_i(\omega), s_{-i}(\omega)) p(\omega).
\end{equation}
When these conditions are met, $s(\omega)$ is the final distribution in equilibrium, which means that it is the same distribution as the correlation probabilities, that the players always follow. Hence this is the correlated equilibrium distribution. The condition in Eq.\ (\ref{conditioned correlation}) states that the players only need to maximize their payoff for the information they possess in a particular moment, without considering alternative information, the latter happening in eq.\ (\ref{correlation}). The conditions that arise from eq.\ (\ref{conditioned correlation}) are the correlated equilibrium conditions. In the above example, we have for player $1$ only that $\bar{s}_1=D$, so the condition is
$$p(CC)\left[u_1(C,C) - u_1(D,C)\right] + p(CD)\left[u_1(C,D) - u_1(D,D)\right] \geq 0.$$

\section{Response Strategy in Game Theory}\label{responsestrategy}

\subsection{Response Probabilities}

We introduce the responses as a new game that the players engage in, this time about playing for or against the advised move. We retain the notion of the correlation device and without loss of generality assume we only have two players, such that the initial correlations are given by $p(\omega')$, with $\omega'\in \{\mu'\nu'\}$ and thus $\mu'$ the available instructions for player $i$, $\nu'$ the available instructions for player $-i$, and both $\mu', \nu' \in \{C,D\}$. In this new game, the "pure" strategies available to the players are to either follow the indications or to not follow them, i.e., $F_{\mu'}$ and $NF_{\mu'}$.  We assign uncorrelated probabilities to the players following or not following the correlation device, $P_{F_{\mu'}}=1-P_{NF_{\mu'}}$, which we call the \textit{response probabilities}. Each instruction that the players can obey creates a new probability variable. A response strategy is defined as $\rho(\omega')=\{P_{F_{\mu'}},P_{F_{\nu'}}\}$. The final distribution of outcomes is represented by a renormalized correlated game with $\omega \in \{\mu \nu\}$ and
\begin{equation}p^R(\omega)=\sum_{\omega'} P_{\mu, \mu'} P_{\nu, \nu'} p(\omega'),
\end{equation}
where the transition probability
\begin{equation}\label{transitions}
P_{\mu, \mu'}= \delta_{\mu, \mu'} P_{F_{\mu'}} + \left(1-\delta_{\mu, \mu'} \right) P_{NF_{\mu'}}
\end{equation} is a function of $P_{F_{\mu'}}$.

With this formulation, we allow for a continuum of reactions from the players to the correlations. The average payoff of player $i$ is given by
\begin{align}
\langle u_i \rangle ^R = \sum_{\omega} u_i(\omega) p^R(\omega) \equiv \sum_{\omega'} u^R_i\left(\omega' \right) p(\omega'),
\end{align}
with
\begin{equation}
u_i^R(\omega') 
=  \sum_{\omega} u_i(\omega) P_{\mu, \mu'} P_{\nu, \nu'}
\end{equation}
and $u_i(\omega)$ as shorthand for $u_i(\text{s}_i(\omega), \text{s}_{-i}(\omega))$.
This result is also equivalent to calculating the convex combination of the uncorrelated games obtained using $P_{F_{\mu'}}$ or $P_{NF_{\mu'}}$ as the probabilities of playing the recommended pure strategy and averaging the payoffs with $p(\omega')$.

\subsection{Slope Analysis in Game Theory}

We can rewrite the slope conditions derived in our paper in formal game-theoretical notation as
\begin{equation} \label{resne}
\sum_{\omega,\nu'}u_i(\omega) P^*_{\mu, \mu'} P^*_{\nu, \nu'} p(\omega') \geq \sum_{\omega,\nu'}u_i(\omega) P_{\mu, \mu'} P^*_{\nu, \nu'} p(\omega'),
\end{equation} where $P^*_{\mu,\mu'}=P_{\mu,\mu'} \left(P^*_{F_{\mu'}} \right)$.
Substituting eq.\ (\ref{transitions}), we get
\begin{align*}
&\sum_{\omega,\nu'} u_i(\omega) \left[ \delta_{\mu,\mu'}P^*_{F_{\mu'}} + (1-\delta_{\mu',\mu})P^*_{NF_{\mu'}} \right] P^*_{\nu, \nu'} p(\omega') \geq \sum_{\omega,\nu'} u_i(\omega) \left[ \delta_{\mu,\mu'}P_{F_{\mu'}} + (1-\delta_{\mu,\mu'})P_{NF_{\mu'}} \right] P^*_{\nu, \nu'} p(\omega'),
\end{align*}
and after performing the sum over $\mu$ we obtain
\begin{align}
& \sum_{\nu, \nu'} P^*_{\nu, \nu'} p(\omega') \left[ \left( P^*_{F_{\mu'}} - P_{F_{\mu'}} \right) u_i(\mu' \nu) +\left( P^*_{NF_{\mu'}} - P_{NF_{\mu'}} \right) u_i(\mu \nu) \right] \geq 0\nonumber \\
& \Leftrightarrow \left( P^*_{F_{\mu'}} - P_{F_{\mu'}} \right) \sum_{\nu, \nu'} P^*_{\nu, \nu'} p(\omega') \left( u_i(\mu' \nu) - u_i(\mu \nu) \right) \geq 0. \label{slopesimple}
\end{align}
The coefficient of $\left( P^*_{F_{\mu'}} - P_{F_{\mu'}} \right)$ is the general form of the slopes that we analyze. Clearly when $ u(\mu \nu) > u(\mu' \nu)$ the overall condition is satisfied if player $i$ follows $\mu'$ with zero probability, which results in a new equilibrium.

Let us assume $\mu'=C$, meaning that the information that player $1$ received was $C$. Substituting in the above, we obtain
\begin{align*}
& \left(P^*_{F_{C_1}}-P_{F_{C_1}}\right) \left( p(CC) \left[ P^*_{F_{C_2}} \left( u_1(CC) - u_1(DC) \right) + P^*_{NF_{C_2}} \left( u_1(CD) - u_1(DD) \right) \right] \right. \\
&+ \left. p(CD) \left[ P^*_{F_{D_2}} \left( u_1(CD) - u_1(DD) \right) + P^*_{NF_{D_2}}  \left( u_1(CC) - u_1(DC) \right) \right] \right) \geq 0.
\end{align*} If the probabilities of following of both players are equal to $1$ we recover the correlated equilibrium condition given as an example in the main part of the paper. This shows that the expected correlated equilibrium is only one of the possible equilibria emerging from the response probabilities.

\subsection{Response Strategy and Nash Equilibrium}

The fact that we do not sum over $\mu'$ in eq.\ \ref{resne} highlights that we have an independent condition for every response probability. Nonetheless, since these probabilities are indeed independent, we can sum over $\mu'$, resulting in
\begin{align*}
& \sum_{\omega, \omega'} u_i\left(s_i(\omega),s_{-i}(\omega)\right) P^*_{\mu, \mu'} P^*_{\nu, \nu'} p(\omega') \geq \sum_{\omega, \omega'} u_i\left(s_i(\omega),s_{-i}(\omega)\right)  P_{\mu, \mu'} P^*_{\nu, \nu'} p(\omega') \\
& \Leftrightarrow \sum_{\omega'} u_i^R\left(\rho^*_i(\omega'),\rho^*_{-i}(\omega') \right) p(\omega') \geq \sum_{\omega'} u_i^R\left(\rho_i(\omega'),\rho^*_{-i}(\omega') \right) p(\omega').
\end{align*}
Both probabilities in $\rho_i(\omega')$, namely $P_{F_{C_i}}$ and $P_{F_{D_i}}$, are independent and each has the same properties as the probability distributions over the pure strategies that correspond to the mixed strategy distributions. They are convex and compact in a finite-dimensional Euclidean space. By the same token as for the mixed strategy, we can be sure to find a fixed point for the response strategies. More specifically, if $P$ is the space of strategy profiles for $P_{F_\mu}$, which thus has a dimensionality equal to the number of possible values of $\mu$, we can define a function called the "reaction correspondence" $r_i$ that maps each response strategy profile $\rho$ to the set of response probabilities that maximize player's $i$ payoff when his opponents play $\rho_{-i}$. The reaction correspondence is defined as $r:P \rightarrow P$. A fixed point of r exists when the players do not have any incentive to change strategy, meaning that they cannot maximize their payoff function any further. The fixed point $\rho^*$ is such that for each player $\rho^*_i \in r_i(\rho^*)$. Thus, a fixed point of $r$ is a response equilibrium, of the same kind as the Nash equilibrium, but now with more probability distributions associated to each player. The proof follows, therefore, analogously from Kakutani's fixed point theorem.

\subsection{Response Strategy and Correlated Equilibrium}

To show that the final game is in correlated equilibrium, we need to make sure that the players actually obey the renormalized probabilities. If the information partitions of player $i$ in two different states are the same, then what player $i$ has played in that state is the same as what he played in the other, which means that having $h_{i}(\omega^a) = h_{i}(\omega^b)$ for two otherwise different states $\omega^a$ and $\omega^b$, is equivalent to having $s_i(\omega^a)=s_i(\omega^b)$. The transition probabilities then describe a mapping from following the initial correlation device to following the renormalized correlation device:
\begin{equation}\label{transitions2}
P_{h_i,h'_i}=\delta_{h_i,h'_i}P_{F_{h'_i}} + (1-\delta_{h_i,h'_i})P_{NF_{h'_i}}.
\end{equation}
where we abbreviate $h_{i}(\omega')$ to $h'_{i}$ and $h_{i}(\omega)$ to $h_{i}$. We can rewrite eq.\ (\ref{resne}) as
\begin{align} \label{resne2}
\sum_{\omega,\{\omega'|h'_i\}} u_i\left(s_i\left(\omega \right),s_{-i}\left( \omega \right) \right)  (P^*_{h_i,h'_i}-P_{h_i,h'_i}) P^*_{h_{-i},h'_{-i}} p(\omega'|h'_i)
\geq 0.
\end{align}
Introducing eq. (\ref{transitions2}) in eq. (\ref{resne2}) we get
\begin{align*}
\sum_{\omega,\{\omega'|h'_i\}} & u_i\left(s_i\left(\omega \right),s_{-i}\left( \omega \right) \right)  \left(\delta_{h_i,h'_i}\left(P^*_{F_{h'_i}}-P_{F_{h'_i}}\right) + (1-\delta_{h_i,h'_i})\left(P^*_{NF_{h'_i}}-P_{NF_{h'_i}} \right) \right) P^*_{h_{-i},h'_{-i}} p(\omega'|h'_i) \geq 0.
\end{align*}
Now we realize that the sum over $\omega$ is equivalent to a sum over $h_i$ and $h_{-i}$. Summing only over $h_i$, we make the transition probabilities effectively act on the payoff functions as
\begin{align*}
\sum_{h_{-i},\{\omega'|h'_i\}} & \left(\underbrace{\left(P^*_{F_{h'_i}}-P_{F_{h'_i}}\right)u_i\left(s_i\left(\omega' \right),s_{-i}\left( \omega \right) \right)}_{h_i=h'_i} \right.\\
&\left.+ \underbrace{\left(P^*_{NF_{h'_i}}-P_{NF_{h'_i}} \right) u_i\left(s_i\left(\bar{\omega'} \right),s_{-i}\left( \omega \right) \right)}_{h_i\neq h'_i} \right) P^*_{h_{-i},h'_{-i}} p(\omega'|h'_i)
\geq 0.
\end{align*}
Simplifying, we arrive at an equation analogous to eq. (\ref{slopesimple}), namely
\begin{align}
&\left(P^*_{F_{h'_i}}-P_{F_{h'_i}}\right) \sum_{h_{-i},\{\omega'|h'_i\}}  \left[u_i\left(s_i\left(\omega' \right),s_{-i}\left( \omega \right) \right) - u_i\left(s_i\left(\bar{\omega'} \right),s_{-i}\left( \omega \right) \right) \right]
P^*_{h_{-i},h'_{-i}} p(\omega'|h'_i)
\geq 0.
\end{align}

We now turn our attention to the coefficients of $\left(P^*_{F_{h'_i}}-P_{F_{h'_i}}\right)$.  Looking at the extreme values of $P^*_{F_{h'_i}}$, we can make the following observations:
\begin{itemize}
\item $P^*_{F_{h'_i}}=1$: $\left(P^*_{F_{h'_i}}-P_{F_{h'_i}}\right)$ is non-negative, such that its coefficient must be positive. In this case, $P^*_{h_i,h'_i}=\delta_{h_i,h'_i}$, which means that we can multiply by this factor and reinstate the sum over $h_i$ and substitute $\omega'$ by $\omega$ and $\bar{\omega'}$ by $\bar{\omega}$;
\item $P^*_{F_{h'_i}}=0$: $\left(P^*_{F_{h'_i}}-P_{F_{h'_i}}\right)$ is non-positive, such that its coefficient must be negative. To have the sum be positive again, we swap the signs of the payoff terms. Multiplying by $P^*_{h_i,h'_i}=1-\delta_{h_i,h'_i}$ and again summing over $h_i$, we have the same expression as in the previous case, as now we need to substitute $\bar{\omega'}$ by $\omega$ and $\omega'$ by $\bar{\omega}$;
\item $0 < P^*_{F_{h'_i}} < 1$:  $\left(P^*_{F_{h'_i}}-P_{F_{h'_i}}\right)$ can be either negative, positive or zero, which means that its coefficient has to be equal to zero. Since that is the case, it is irrelevant which sign the payoff terms have, and so we can multiply by $P^*_{h_i,h'_i}$ and sum over $h_i$ freely. We then arrive again at the same formula, if we make either one of the substitutions of the previous cases, but now equating to zero.
\end{itemize}

With this, we can thus rewrite the slope as
\begin{align}
\sum_{\omega,\{\omega'|h'_i\}}  P^*_{h_{i},h'_{i}}  P^*_{h_{-i},h'_{-i}} p(\omega'|h'_i)  \left[u_i\left(s_i\left(\omega \right),s_{-i}\left( \omega \right) \right) - u_i\left(s_i\left(\bar{\omega} \right),s_{-i}\left( \omega \right)  \right)\right]
\geq 0,
\end{align} for any $\bar{\omega}$.
Because the sum is not over $h'_i$, the product of probabilities is equivalent to the conditioning of the renormalized probability distribution on $h'_i$, which leads to
\begin{align}
\sum_{\{\omega|h'_i\}} p^R(\omega|h'_i)  \left[u_i\left(s_i\left(\omega \right),s_{-i}\left( \omega \right) \right) - u_i\left(s_i\left(\bar{\omega} \right),s_{-i}\left( \omega \right)  \right)\right]
\geq 0,
\end{align} 
such that if the response strategies are in Nash equilibrium, then the final distribution is Bayes rational towards the state of the world $\omega$. Multiplying by $P_{h'_i}$ and summing over $h'_i$ gives
\begin{align}
\sum_{\omega} p^R(\omega) \left[u_i\left(s_i\left(\omega \right),s_{-i}\left( \omega \right) \right) - u_i\left(s_i\left(\bar{\omega} \right),s_{-i}\left( \omega \right) \right)\right] 
\geq 0.
\end{align}
If the target final distribution is the same as that of $s(\omega)$, then the action corresponding to playing accordingly to a different final distribution is equivalent to playing a different action in $S_i$ with respect to the final distribution, such that $s_i\left(\bar{\omega} \right) = \bar{s}_i(\omega)$, and we arrive at
\begin{align}
\sum_{\omega} p^R(\omega) \left[u_i\left(s_i\left(\omega \right),s_{-i}\left( \omega \right) \right) - u_i\left(\bar{s}_i(\omega),s_{-i}\left( \omega \right)  \right)\right]
\geq 0,
\end{align} 
which is a game with a renormalized distribution that is in correlated equilibrium. With this we can conclude that if the response strategies are in equilibrium, then the final distribution is a correlated equilibrium.

The players are Bayes rational towards the initial world when they both want to follow, since $\omega= \omega'$. When that is not the case, the response probabilities allow them to find a world towards which they want to be rational. Thus we see that the response probabilities create a condition for which the main theorem in \cite{aumann1987correlated} applies.
\\

\bigskip

\twocolumngrid

\begin{acknowledgments}
We thank Joris Broere and Vincent Buskens for discussions during the research. This work is supported by the Complex Systems Fund, with special thanks to Peter Koeze.
\end{acknowledgments}

\bibliography{bib}

\providecommand{\noopsort}[1]{}\providecommand{\singleletter}[1]{#1}%
\begin{thebibliography}{45}%
\makeatletter
\providecommand \@ifxundefined [1]{%
 \@ifx{#1\undefined}
}%
\providecommand \@ifnum [1]{%
 \ifnum #1\expandafter \@firstoftwo
 \else \expandafter \@secondoftwo
 \fi
}%
\providecommand \@ifx [1]{%
 \ifx #1\expandafter \@firstoftwo
 \else \expandafter \@secondoftwo
 \fi
}%
\providecommand \natexlab [1]{#1}%
\providecommand \enquote  [1]{``#1''}%
\providecommand \bibnamefont  [1]{#1}%
\providecommand \bibfnamefont [1]{#1}%
\providecommand \citenamefont [1]{#1}%
\providecommand \href@noop [0]{\@secondoftwo}%
\providecommand \href [0]{\begingroup \@sanitize@url \@href}%
\providecommand \@href[1]{\@@startlink{#1}\@@href}%
\providecommand \@@href[1]{\endgroup#1\@@endlink}%
\providecommand \@sanitize@url [0]{\catcode `\\12\catcode `\$12\catcode
  `\&12\catcode `\#12\catcode `\^12\catcode `\_12\catcode `\%12\relax}%
\providecommand \@@startlink[1]{}%
\providecommand \@@endlink[0]{}%
\providecommand \url  [0]{\begingroup\@sanitize@url \@url }%
\providecommand \@url [1]{\endgroup\@href {#1}{\urlprefix }}%
\providecommand \urlprefix  [0]{URL }%
\providecommand \Eprint [0]{\href }%
\providecommand \doibase [0]{http://dx.doi.org/}%
\providecommand \selectlanguage [0]{\@gobble}%
\providecommand \bibinfo  [0]{\@secondoftwo}%
\providecommand \bibfield  [0]{\@secondoftwo}%
\providecommand \translation [1]{[#1]}%
\providecommand \BibitemOpen [0]{}%
\providecommand \bibitemStop [0]{}%
\providecommand \bibitemNoStop [0]{.\EOS\space}%
\providecommand \EOS [0]{\spacefactor3000\relax}%
\providecommand \BibitemShut  [1]{\csname bibitem#1\endcsname}%
\let\auto@bib@innerbib\@empty
\bibitem [{\citenamefont {Feynman}(1954)}]{feyn54}%
  \BibitemOpen
  \bibfield  {author} {\bibinfo {author} {\bibfnamefont {R.~P.}\ \bibnamefont
  {Feynman}},\ }\href@noop {} {\bibfield  {journal} {\bibinfo  {journal}
  {Phys.\ Rev.}\ }\textbf {\bibinfo {volume} {94}},\ \bibinfo {pages} {262}
  (\bibinfo {year} {1954})}\BibitemShut {NoStop}%
\bibitem [{\citenamefont {Witten}(2001)}]{witten2001}%
  \BibitemOpen
  \bibfield  {author} {\bibinfo {author} {\bibfnamefont {E.}~\bibnamefont
  {Witten}},\ }\href@noop {} {} (\bibinfo {year} {2001}),\ \Eprint
  {http://arxiv.org/abs/hep-th/0106109} {hep-th/0106109} \BibitemShut {NoStop}%
\bibitem [{\citenamefont {Einstein}\ \emph {et~al.}(1935)\citenamefont
  {Einstein}, \citenamefont {Podolsky},\ and\ \citenamefont {Rosen}}]{epr}%
  \BibitemOpen
  \bibfield  {author} {\bibinfo {author} {\bibfnamefont {A.}~\bibnamefont
  {Einstein}}, \bibinfo {author} {\bibfnamefont {{\relax Yu}.}~\bibnamefont
  {Podolsky}}, \ and\ \bibinfo {author} {\bibfnamefont {N.}~\bibnamefont
  {Rosen}},\ }\href@noop {} {\bibfield  {journal} {\bibinfo  {journal} {Phys.\
  Rev.}\ }\textbf {\bibinfo {volume} {47}},\ \bibinfo {pages} {777} (\bibinfo
  {year} {1935})}\BibitemShut {NoStop}%
\bibitem [{\citenamefont {Birell}\ and\ \citenamefont {Davies}(1982)}]{Bire82}%
  \BibitemOpen
  \bibfield  {author} {\bibinfo {author} {\bibfnamefont {N.~D.}\ \bibnamefont
  {Birell}}\ and\ \bibinfo {author} {\bibfnamefont {P.~C.~W.}\ \bibnamefont
  {Davies}},\ }\href@noop {} {\emph {\bibinfo {title} {Quantum Fields in Curved
  Space}}}\ (\bibinfo  {publisher} {Cambridge University Press},\ \bibinfo
  {year} {1982})\BibitemShut {NoStop}%
\bibitem [{\citenamefont {Berman}\ and\ \citenamefont
  {Izrailev}(1983)}]{Berman1983}%
  \BibitemOpen
  \bibfield  {author} {\bibinfo {author} {\bibfnamefont {G.~P.}\ \bibnamefont
  {Berman}, \bibfnamefont {Jr.}}\ and\ \bibinfo {author} {\bibfnamefont
  {F.~M.}\ \bibnamefont {Izrailev}, \bibfnamefont {Jr.}},\ }\href@noop {}
  {\bibfield  {journal} {\bibinfo  {journal} {Physica D}\ }\textbf {\bibinfo
  {volume} {88}},\ \bibinfo {pages} {445} (\bibinfo {year} {1983})}\BibitemShut
  {NoStop}%
\bibitem [{Note1()}]{Note1}%
  \BibitemOpen
  \bibinfo {note} {Automatically placing footnotes into the bibliography
  requires using BibTeX to compile the bibliography.}\BibitemShut {Stop}%
\bibitem [{\citenamefont {Davies}\ and\ \citenamefont
  {Parns}(1988)}]{Davies1998}%
  \BibitemOpen
  \bibfield  {author} {\bibinfo {author} {\bibfnamefont {E.~B.}\ \bibnamefont
  {Davies}}\ and\ \bibinfo {author} {\bibfnamefont {L.}~\bibnamefont {Parns}},\
  }\href@noop {} {\bibfield  {journal} {\bibinfo  {journal} {Q. J. Mech. Appl.
  Math.}\ }\textbf {\bibinfo {volume} {51}},\ \bibinfo {pages} {477} (\bibinfo
  {year} {1988})}\BibitemShut {NoStop}%
\bibitem [{\citenamefont {Beutler}(1994{\natexlab{a}})}]{Beutler1994}%
  \BibitemOpen
  \bibfield  {author} {\bibinfo {author} {\bibfnamefont {E.}~\bibnamefont
  {Beutler}},\ }\enquote {\bibinfo {title} {Williams hematology},}\ \ (\bibinfo
   {publisher} {McGraw-Hill},\ \bibinfo {address} {New York},\ \bibinfo {year}
  {1994})\ Chap.~\bibinfo {chapter} {7}, pp.\ \bibinfo {pages} {654--662},\
  \bibinfo {edition} {5th}\ ed.\BibitemShut {Stop}%
\bibitem [{\citenamefont {Knuth}(1973)}]{inbook-full}%
  \BibitemOpen
  \bibfield  {author} {\bibinfo {author} {\bibfnamefont {D.~E.}\ \bibnamefont
  {Knuth}},\ }\enquote {\bibinfo {title} {Fundamental algorithms},}\ \
  (\bibinfo  {publisher} {Addison-Wesley},\ \bibinfo {address} {Reading,
  Massachusetts},\ \bibinfo {year} {\noopsort{1973b}1973})\ \bibinfo {type}
  {Section}\ \bibinfo {chapter} {1.2}, pp.\ \bibinfo {pages} {10--119},\
  \bibinfo {edition} {2nd}\ ed.,\ \bibinfo {note} {a full INBOOK
  entry}\BibitemShut {NoStop}%
\bibitem [{\citenamefont {Smith}\ and\ \citenamefont
  {Johnson}(2005)}]{Smith2005}%
  \BibitemOpen
  \bibfield  {author} {\bibinfo {author} {\bibfnamefont {J.~S.}\ \bibnamefont
  {Smith}}\ and\ \bibinfo {author} {\bibfnamefont {G.~W.}\ \bibnamefont
  {Johnson}},\ }\href@noop {} {\bibfield  {journal} {\bibinfo  {journal}
  {Philos. Trans. R. Soc. London, Ser. B}\ }\textbf {\bibinfo {volume} {777}},\
  \bibinfo {pages} {1395} (\bibinfo {year} {2005})}\BibitemShut {NoStop}%
\bibitem [{\citenamefont {Smith}\ \emph {et~al.}({\natexlab{a}})\citenamefont
  {Smith}, \citenamefont {Johnson},\ and\ \citenamefont {Miller}}]{Smith2010}%
  \BibitemOpen
  \bibfield  {author} {\bibinfo {author} {\bibfnamefont {W.~J.}\ \bibnamefont
  {Smith}}, \bibinfo {author} {\bibfnamefont {T.~J.}\ \bibnamefont {Johnson}},
  \ and\ \bibinfo {author} {\bibfnamefont {B.~G.}\ \bibnamefont {Miller}},\
  }\href@noop {} {\enquote {\bibinfo {title} {Surface chemistry and
  preferential crystal orientation on a silicon surface},}\ }
  ({\natexlab{a}}),\ \bibinfo {note} {{J. Appl. Phys.}
  (unpublished)}\BibitemShut {NoStop}%
\bibitem [{\citenamefont {Smith}\ \emph {et~al.}({\natexlab{b}})\citenamefont
  {Smith}, \citenamefont {Johnson},\ and\ \citenamefont {Klein}}]{Smith2010a}%
  \BibitemOpen
  \bibfield  {author} {\bibinfo {author} {\bibfnamefont {V.~K.}\ \bibnamefont
  {Smith}}, \bibinfo {author} {\bibfnamefont {K.}~\bibnamefont {Johnson}}, \
  and\ \bibinfo {author} {\bibfnamefont {M.~O.}\ \bibnamefont {Klein}},\
  }\href@noop {} {\enquote {\bibinfo {title} {Surface chemistry and
  preferential crystal orientation on a silicon surface},}\ }
  ({\natexlab{b}}),\ \bibinfo {note} {{J. Appl. Phys.} (submitted)}\BibitemShut
  {NoStop}%
\bibitem [{\citenamefont {{\"{U}}nderwood}\ \emph {et~al.}(1988)\citenamefont
  {{\"{U}}nderwood}, \citenamefont {{\~N}et},\ and\ \citenamefont
  {{\={P}}ot}}]{unpublished-full}%
  \BibitemOpen
  \bibfield  {author} {\bibinfo {author} {\bibfnamefont {U.}~\bibnamefont
  {{\"{U}}nderwood}}, \bibinfo {author} {\bibfnamefont {N.}~\bibnamefont
  {{\~N}et}}, \ and\ \bibinfo {author} {\bibfnamefont {P.}~\bibnamefont
  {{\={P}}ot}},\ }\href@noop {} {\enquote {\bibinfo {title} {Lower bounds for
  wishful research results},}\ } (\bibinfo {year} {1988}),\ \bibinfo {note}
  {talk at Fanstord University (A full UNPUBLISHED entry)}\BibitemShut
  {NoStop}%
\bibitem [{\citenamefont {Johnson}\ \emph {et~al.}(2007)\citenamefont
  {Johnson}, \citenamefont {Miller},\ and\ \citenamefont
  {Smith}}]{JohnsonMillerSmith2007}%
  \BibitemOpen
  \bibfield  {author} {\bibinfo {author} {\bibfnamefont {M.~P.}\ \bibnamefont
  {Johnson}}, \bibinfo {author} {\bibfnamefont {K.~L.}\ \bibnamefont {Miller}},
  \ and\ \bibinfo {author} {\bibfnamefont {K.}~\bibnamefont {Smith}},\
  }\href@noop {} {}\bibinfo {howpublished} {personal communication} (\bibinfo
  {year} {2007})\BibitemShut {NoStop}%
\bibitem [{\citenamefont {Smith}(2007{\natexlab{a}})}]{Smith2007}%
  \BibitemOpen
  \bibinfo {editor} {\bibfnamefont {J.}~\bibnamefont {Smith}},\ ed.,\
  \href@noop {} {\emph {\bibinfo {title} {AIP Conf. Proc.}}},\ Vol.\ \bibinfo
  {volume} {841}\ (\bibinfo {year} {2007})\BibitemShut {NoStop}%
\bibitem [{\citenamefont {Oz}\ and\ \citenamefont
  {Yannakakis}(1983)}]{proceedings-full}%
  \BibitemOpen
  \bibinfo {editor} {\bibfnamefont {W.~V.}\ \bibnamefont {Oz}}\ and\ \bibinfo
  {editor} {\bibfnamefont {M.}~\bibnamefont {Yannakakis}},\ eds.,\ \href@noop
  {} {\emph {\bibinfo {title} {Proc. Fifteenth Annual}}},\ \bibinfo {series}
  {All ACM Conferences}\ No.~\bibinfo {number} {17},\ \bibinfo {organization}
  {ACM}\ (\bibinfo  {publisher} {Academic Press},\ \bibinfo {address}
  {Boston},\ \bibinfo {year} {1983})\ \bibinfo {note} {a full PROCEEDINGS
  entry}\BibitemShut {NoStop}%
\bibitem [{\citenamefont {Burstyn}(2004)}]{Burstyn2004}%
  \BibitemOpen
  \bibfield  {author} {\bibinfo {author} {\bibfnamefont {Y.}~\bibnamefont
  {Burstyn}},\ }\href@noop {} {\enquote {\bibinfo {title} {{Proceedings of the
  5th International Molecular Beam Epitaxy Conference, Santa Fe, NM}},}\ }
  (\bibinfo {year} {2004}),\ \bibinfo {note} {(unpublished)}\BibitemShut
  {NoStop}%
\bibitem [{\citenamefont {Quinn}(2001)}]{Quinn2001}%
  \BibitemOpen
  \bibinfo {editor} {\bibfnamefont {B.}~\bibnamefont {Quinn}},\ ed.,\
  \href@noop {} {\emph {\bibinfo {title} {{Proceedings of the 2003 Particle
  Accelerator Conference, Portland, OR, 12-16 May 2005}}}}\ (\bibinfo
  {publisher} {Wiley},\ \bibinfo {address} {New York},\ \bibinfo {year}
  {2001})\ \bibinfo {note} {albeit the conference was held in 2005, it was the
  2003 conference, and the proceedings were published in 2001; go
  figure}\BibitemShut {NoStop}%
\bibitem [{\citenamefont {Agarwal}(2001)}]{Agarwal2001}%
  \BibitemOpen
  \bibfield  {author} {\bibinfo {author} {\bibfnamefont {A.~G.}\ \bibnamefont
  {Agarwal}},\ }\href@noop {} {\bibfield  {journal} {\bibinfo  {journal}
  {Semiconductors}\ }\textbf {\bibinfo {volume} {66}},\ \bibinfo {pages} {1238}
  (\bibinfo {year} {2001})}\BibitemShut {NoStop}%
\bibitem [{\citenamefont {Smith}()}]{SmithDA01}%
  \BibitemOpen
  \bibfield  {author} {\bibinfo {author} {\bibfnamefont {R.}~\bibnamefont
  {Smith}},\ }\href@noop {} {\bibfield  {journal} {\bibinfo  {journal} {J.
  Appl. Phys. (these proceedings)}\ }}\bibinfo {note} {Abstract No.
  DA-01}\BibitemShut {NoStop}%
\bibitem [{\citenamefont {Smith}(2007{\natexlab{b}})}]{Smith2007a}%
  \BibitemOpen
  \bibfield  {author} {\bibinfo {author} {\bibfnamefont {J.}~\bibnamefont
  {Smith}},\ }\href@noop {} {\bibfield  {journal} {\bibinfo  {journal} {Proc.
  SPIE}\ }\textbf {\bibinfo {volume} {124}},\ \bibinfo {pages} {367} (\bibinfo
  {year} {2007}{\natexlab{b}})},\ \bibinfo {note} {required title is
  missing}\BibitemShut {NoStop}%
\bibitem [{\citenamefont {T{\'{e}}rrific}(1988)}]{techreport-full}%
  \BibitemOpen
  \bibfield  {author} {\bibinfo {author} {\bibfnamefont {T.}~\bibnamefont
  {T{\'{e}}rrific}},\ }\href@noop {} {\emph {\bibinfo {title} {An {$O(n \log n
  / \! \log\log n)$} Sorting Algorithm}}},\ \bibinfo {type} {Wishful Research
  Result}\ \bibinfo {number} {7}\ (\bibinfo  {institution} {Fanstord
  University},\ \bibinfo {address} {Computer Science Department, Fanstord,
  California},\ \bibinfo {year} {1988})\ \bibinfo {note} {a full TECHREPORT
  entry}\BibitemShut {NoStop}%
\bibitem [{\citenamefont {Nelson}(1999{\natexlab{a}})}]{Nelson1999}%
  \BibitemOpen
  \bibfield  {author} {\bibinfo {author} {\bibfnamefont {J.}~\bibnamefont
  {Nelson}},\ }\href@noop {} {}\bibinfo {type} {{TWI Report}}\ \bibinfo
  {number} {666/1999}\ (\bibinfo {year} {Jan.~1999})\ \bibinfo {note} {required
  institution missing}\BibitemShut {NoStop}%
\bibitem [{\citenamefont {Fields}(2005)}]{Fields2005}%
  \BibitemOpen
  \bibfield  {author} {\bibinfo {author} {\bibfnamefont {W.~K.}\ \bibnamefont
  {Fields}},\ }\href@noop {} {}\bibinfo {type} {{ECE Report No.}}\ \bibinfo
  {number} {AL944}\ (\bibinfo {year} {2005})\ \bibinfo {note} {required
  institution missing}\BibitemShut {NoStop}%
\bibitem [{\citenamefont {Zalkins}(2008)}]{Zalkins2008}%
  \BibitemOpen
  \bibfield  {author} {\bibinfo {author} {\bibfnamefont {Y.~M.}\ \bibnamefont
  {Zalkins}},\ }\href@noop {} {}\bibinfo {howpublished} {e-print
  arXiv:cond-mat/040426} (\bibinfo {year} {2008})\BibitemShut {NoStop}%
\bibitem [{\citenamefont {Nelson}(2005)}]{Nelson2005}%
  \BibitemOpen
  \bibfield  {author} {\bibinfo {author} {\bibfnamefont {J.}~\bibnamefont
  {Nelson}},\ }\href@noop {} {}\bibinfo {howpublished} {{U.S. Patent No.}
  5,693,000} (\bibinfo {year} {12~Dec.~2005})\BibitemShut {NoStop}%
\bibitem [{\citenamefont {Nelson}(1999{\natexlab{b}})}]{Nelson1999a}%
  \BibitemOpen
  \bibfield  {author} {\bibinfo {author} {\bibfnamefont {J.~K.}\ \bibnamefont
  {Nelson}},\ }\href@noop {} {\bibinfo {type} {M.{S}. thesis}},\ \bibinfo
  {school} {New York University} (\bibinfo {year}
  {1999}{\natexlab{b}})\BibitemShut {NoStop}%
\bibitem [{\citenamefont {Masterly}(1988)}]{mastersthesis-full}%
  \BibitemOpen
  \bibfield  {author} {\bibinfo {author} {\bibfnamefont {{\'{E}}.}~\bibnamefont
  {Masterly}},\ }\emph {\bibinfo {title} {Mastering Thesis Writing}},\
  \href@noop {} {\bibinfo {type} {Master's project}},\ \bibinfo  {school}
  {Stanford University}, \bibinfo {address} {English Department} (\bibinfo
  {year} {1988}),\ \bibinfo {note} {a full MASTERSTHESIS entry}\BibitemShut
  {NoStop}%
\bibitem [{\citenamefont {Smith}(2003)}]{Smith2003}%
  \BibitemOpen
  \bibfield  {author} {\bibinfo {author} {\bibfnamefont {S.~M.}\ \bibnamefont
  {Smith}},\ }\href@noop {} {\bibinfo {type} {{Ph.D.} thesis}},\ \bibinfo
  {school} {Massachusetts Institute of Technology} (\bibinfo {year}
  {2003})\BibitemShut {NoStop}%
\bibitem [{\citenamefont {Kawa}\ and\ \citenamefont {Lin}(2003)}]{KawaLin2003}%
  \BibitemOpen
  \bibfield  {author} {\bibinfo {author} {\bibfnamefont {S.~R.}\ \bibnamefont
  {Kawa}}\ and\ \bibinfo {author} {\bibfnamefont {S.-J.}\ \bibnamefont {Lin}},\
  }\href@noop {} {\bibfield  {journal} {\bibinfo  {journal} {J. Geophys. Res.}\
  }\textbf {\bibinfo {volume} {108}},\ \bibinfo {pages} {4201} (\bibinfo {year}
  {2003})},\ \bibinfo {note} {{DOI:10.1029/2002JD002268}}\BibitemShut {NoStop}%
\bibitem [{\citenamefont {Phony-Baloney}(1988)}]{phdthesis-full}%
  \BibitemOpen
  \bibfield  {author} {\bibinfo {author} {\bibfnamefont {F.~P.}\ \bibnamefont
  {Phony-Baloney}},\ }\emph {\bibinfo {title} {Fighting Fire with Fire:
  Festooning {F}rench Phrases}},\ \href@noop {} {\bibinfo {type} {{PhD}
  dissertation}},\ \bibinfo  {school} {Fanstord University}, \bibinfo {address}
  {Department of French} (\bibinfo {year} {1988}),\ \bibinfo {note} {a full
  PHDTHESIS entry}\BibitemShut {NoStop}%
\bibitem [{\citenamefont {Knuth}(1981)}]{book-full}%
  \BibitemOpen
  \bibfield  {author} {\bibinfo {author} {\bibfnamefont {D.~E.}\ \bibnamefont
  {Knuth}},\ }\href@noop {} {\emph {\bibinfo {title} {Seminumerical
  Algorithms}}},\ \bibinfo {edition} {2nd}\ ed.,\ \bibinfo {series} {The Art of
  Computer Programming}, Vol.~\bibinfo {volume} {2}\ (\bibinfo  {publisher}
  {Addison-Wesley},\ \bibinfo {address} {Reading, Massachusetts},\ \bibinfo
  {year} {\noopsort{1973c}1981})\ \bibinfo {note} {a full BOOK
  entry}\BibitemShut {NoStop}%
\bibitem [{\citenamefont {Knvth}(1988)}]{booklet-full}%
  \BibitemOpen
  \bibfield  {author} {\bibinfo {author} {\bibfnamefont {J.~C.}\ \bibnamefont
  {Knvth}},\ }\href@noop {} {\enquote {\bibinfo {title} {The programming of
  computer art},}\ }\bibinfo {howpublished} {Vernier Art Center},\ \bibinfo
  {address} {Stanford, California} (\bibinfo {year} {1988}),\ \bibinfo {note}
  {a full BOOKLET entry}\BibitemShut {NoStop}%
\bibitem [{\citenamefont {Ballagh}\ and\ \citenamefont
  {Savage}(2000{\natexlab{a}})}]{ballagh2000}%
  \BibitemOpen
  \bibfield  {author} {\bibinfo {author} {\bibfnamefont {R.}~\bibnamefont
  {Ballagh}}\ and\ \bibinfo {author} {\bibfnamefont {C.}~\bibnamefont
  {Savage}},\ }\enquote {\bibinfo {title} {Bose-einstein condensation: from
  atomic physics to quantum fluids, proceedings of the 13th physics summer
  school},}\ \ (\bibinfo  {publisher} {World Scientific},\ \bibinfo {address}
  {Singapore},\ \bibinfo {year} {2000})\ \Eprint
  {http://arxiv.org/abs/cond-mat/0008070} {cond-mat/0008070} \BibitemShut
  {NoStop}%
\bibitem [{\citenamefont {Opechowski}\ and\ \citenamefont
  {Guccione}()}]{Magnetism}%
  \BibitemOpen
  \bibfield  {author} {\bibinfo {author} {\bibfnamefont {W.}~\bibnamefont
  {Opechowski}}\ and\ \bibinfo {author} {\bibfnamefont {R.}~\bibnamefont
  {Guccione}},\ }\enquote {\bibinfo {title} {Introduction to the theory of
  normal metals},}\ in\ \href@noop {} {\emph {\bibinfo {booktitle}
  {Magnetism}}},\ Vol.\ \bibinfo {volume} {IIa},\ \bibinfo {editor} {edited by\
  \bibinfo {editor} {\bibfnamefont {G.~T.}\ \bibnamefont {Rado}}\ and\ \bibinfo
  {editor} {\bibfnamefont {H.}~\bibnamefont {Suhl}}}\ (\bibinfo  {publisher}
  {Academic Press},\ \bibinfo {address} {New York})\ p.\ \bibinfo {pages}
  {105}\BibitemShut {NoStop}%
\bibitem [{\citenamefont {Opechowski}\ and\ \citenamefont
  {Guccione}(1965{\natexlab{a}})}]{Magnetismb}%
  \BibitemOpen
  \bibfield  {author} {\bibinfo {author} {\bibfnamefont {W.}~\bibnamefont
  {Opechowski}}\ and\ \bibinfo {author} {\bibfnamefont {R.}~\bibnamefont
  {Guccione}},\ }in\ \href@noop {} {\emph {\bibinfo {booktitle} {Magnetism}}},\
  Vol.\ \bibinfo {volume} {IIa},\ \bibinfo {editor} {edited by\ \bibinfo
  {editor} {\bibfnamefont {G.~T.}\ \bibnamefont {Rado}}\ and\ \bibinfo {editor}
  {\bibfnamefont {H.}~\bibnamefont {Suhl}}}\ (\bibinfo  {publisher} {Academic
  Press},\ \bibinfo {address} {New York},\ \bibinfo {year} {1965})\ p.\
  \bibinfo {pages} {105}\BibitemShut {NoStop}%
\bibitem [{\citenamefont {Smith}(1980{\natexlab{a}})}]{Smith80}%
  \BibitemOpen
  \bibfield  {author} {\bibinfo {author} {\bibfnamefont {J.~M.}\ \bibnamefont
  {Smith}},\ }\enquote {\bibinfo {title} {Molecular dynamics},}\ \ (\bibinfo
  {publisher} {Academic},\ \bibinfo {address} {New York},\ \bibinfo {year}
  {1980})\BibitemShut {NoStop}%
\bibitem [{\citenamefont {Zakharov}\ and\ \citenamefont {Shabat}(1971)}]{ZS71}%
  \BibitemOpen
  \bibfield  {author} {\bibinfo {author} {\bibfnamefont {V.~E.}\ \bibnamefont
  {Zakharov}}\ and\ \bibinfo {author} {\bibfnamefont {A.~B.}\ \bibnamefont
  {Shabat}},\ }\href@noop {} {\bibfield  {journal} {\bibinfo  {journal} {Zh.
  Eksp. Teor. Fiz.}\ }\textbf {\bibinfo {volume} {61}},\ \bibinfo {pages} {118}
  (\bibinfo {year} {1971})},\ \translation{Sov. Phys. JETP \textbf{34}, 62
  (1972)}\BibitemShut {NoStop}%
\bibitem [{\citenamefont {Beutler}(1994{\natexlab{b}})}]{Beutler1994a}%
  \BibitemOpen
  \bibfield  {author} {\bibinfo {author} {\bibfnamefont {E.}~\bibnamefont
  {Beutler}},\ }in\ \href@noop {} {\emph {\bibinfo {booktitle} {Williams
  Hematology}}},\ Vol.~\bibinfo {volume} {2},\ \bibinfo {editor} {edited by\
  \bibinfo {editor} {\bibfnamefont {E.}~\bibnamefont {Beutler}}, \bibinfo
  {editor} {\bibfnamefont {M.~A.}\ \bibnamefont {Lichtman}}, \bibinfo {editor}
  {\bibfnamefont {B.~W.}\ \bibnamefont {Coller}}, \ and\ \bibinfo {editor}
  {\bibfnamefont {T.~S.}\ \bibnamefont {Kipps}}}\ (\bibinfo  {publisher}
  {McGraw-Hill},\ \bibinfo {address} {New York},\ \bibinfo {year} {1994})\
  \bibinfo {edition} {5th}\ ed.,\ Chap.~\bibinfo {chapter} {7}, pp.\ \bibinfo
  {pages} {654--662}\BibitemShut {NoStop}%
\bibitem [{\citenamefont {Ballagh}\ and\ \citenamefont
  {Savage}(2000{\natexlab{b}})}]{ballagh2000a}%
  \BibitemOpen
  \bibfield  {author} {\bibinfo {author} {\bibfnamefont {R.}~\bibnamefont
  {Ballagh}}\ and\ \bibinfo {author} {\bibfnamefont {C.}~\bibnamefont
  {Savage}},\ }in\ \href@noop {} {\emph {\bibinfo {booktitle} {Proceedings of
  the 13th Physics Summer School}}},\ \bibinfo {editor} {edited by\ \bibinfo
  {editor} {\bibfnamefont {C.}~\bibnamefont {Savage}}\ and\ \bibinfo {editor}
  {\bibfnamefont {M.}~\bibnamefont {Das}}}\ (\bibinfo  {publisher} {World
  Scientific},\ \bibinfo {address} {Singapore},\ \bibinfo {year} {2000})\
  \Eprint {http://arxiv.org/abs/cond-mat/0008070} {cond-mat/0008070}
  \BibitemShut {NoStop}%
\bibitem [{\citenamefont {Opechowski}\ and\ \citenamefont
  {Guccione}(1965{\natexlab{b}})}]{Magnetisma}%
  \BibitemOpen
  \bibfield  {author} {\bibinfo {author} {\bibfnamefont {W.}~\bibnamefont
  {Opechowski}}\ and\ \bibinfo {author} {\bibfnamefont {R.}~\bibnamefont
  {Guccione}},\ }in\ \href@noop {} {\emph {\bibinfo {booktitle} {Magnetism}}},\
  Vol.\ \bibinfo {volume} {IIa},\ \bibinfo {editor} {edited by\ \bibinfo
  {editor} {\bibfnamefont {G.~T.}\ \bibnamefont {Rado}}\ and\ \bibinfo {editor}
  {\bibfnamefont {H.}~\bibnamefont {Suhl}}}\ (\bibinfo  {publisher} {Academic
  Press},\ \bibinfo {address} {New York},\ \bibinfo {year} {1965})\ p.\
  \bibinfo {pages} {105}\BibitemShut {NoStop}%
\bibitem [{\citenamefont {Smith}(1980{\natexlab{b}})}]{Smith80a}%
  \BibitemOpen
  \bibfield  {author} {\bibinfo {author} {\bibfnamefont {J.~M.}\ \bibnamefont
  {Smith}},\ }in\ \href@noop {} {\emph {\bibinfo {booktitle} {Molecular
  Dynamics}}},\ \bibinfo {editor} {edited by\ \bibinfo {editor} {\bibfnamefont
  {C.}~\bibnamefont {Brown}}}\ (\bibinfo  {publisher} {Academic},\ \bibinfo
  {address} {New York},\ \bibinfo {year} {1980})\BibitemShut {NoStop}%
\bibitem [{\citenamefont {Lincoll}(1977)}]{incollection-full}%
  \BibitemOpen
  \bibfield  {author} {\bibinfo {author} {\bibfnamefont {D.~D.}\ \bibnamefont
  {Lincoll}},\ }in\ \href@noop {} {\emph {\bibinfo {booktitle} {High Speed
  Computer and Algorithm Organization}}},\ \bibinfo {series and number}
  {\bibinfo {series} {Fast Computers}\ No.~\bibinfo {number} {23}},\ \bibinfo
  {editor} {edited by\ \bibinfo {editor} {\bibfnamefont {D.~J.}\ \bibnamefont
  {Lipcoll}}, \bibinfo {editor} {\bibfnamefont {D.~H.}\ \bibnamefont {Lawrie}},
  \ and\ \bibinfo {editor} {\bibfnamefont {A.~H.}\ \bibnamefont {Sameh}}}\
  (\bibinfo  {publisher} {Academic Press},\ \bibinfo {address} {New York},\
  \bibinfo {year} {1977})\ \bibinfo {edition} {3rd}\ ed.,\ \bibinfo {type}
  {Part}~\bibinfo {chapter} {3}, pp.\ \bibinfo {pages} {179--183},\ \bibinfo
  {note} {a full INCOLLECTION entry}\BibitemShut {NoStop}%
\bibitem [{\citenamefont {Oaho}\ \emph {et~al.}(1983)\citenamefont {Oaho},
  \citenamefont {Ullman},\ and\ \citenamefont
  {Yannakakis}}]{inproceedings-full}%
  \BibitemOpen
  \bibfield  {author} {\bibinfo {author} {\bibfnamefont {A.~V.}\ \bibnamefont
  {Oaho}}, \bibinfo {author} {\bibfnamefont {J.~D.}\ \bibnamefont {Ullman}}, \
  and\ \bibinfo {author} {\bibfnamefont {M.}~\bibnamefont {Yannakakis}},\ }in\
  \href@noop {} {\emph {\bibinfo {booktitle} {Proc. Fifteenth Annual ACM}}},\
  \bibinfo {series and number} {\bibinfo {series} {All ACM Conferences}\
  No.~\bibinfo {number} {17}},\ \bibinfo {editor} {edited by\ \bibinfo {editor}
  {\bibfnamefont {W.~V.}\ \bibnamefont {Oz}}\ and\ \bibinfo {editor}
  {\bibfnamefont {M.}~\bibnamefont {Yannakakis}}},\ \bibinfo {organization}
  {ACM}\ (\bibinfo  {publisher} {Academic Press},\ \bibinfo {address}
  {Boston},\ \bibinfo {year} {1983})\ pp.\ \bibinfo {pages} {133--139},\
  \bibinfo {note} {a full INPROCEDINGS entry}\BibitemShut {NoStop}%
\bibitem [{\citenamefont {Manmaker}(1986)}]{manual-full}%
  \BibitemOpen
  \bibfield  {author} {\bibinfo {author} {\bibfnamefont {L.}~\bibnamefont
  {Manmaker}},\ }\href@noop {} {\emph {\bibinfo {title} {The Definitive
  Computer Manual}}},\ \bibinfo {organization} {Chips-R-Us},\ \bibinfo
  {address} {Silicon Valley},\ \bibinfo {edition} {silver}\ ed. (\bibinfo
  {year} {1986}),\ \bibinfo {note} {a full MANUAL entry}\BibitemShut {NoStop}%
\end{thebibliography}%


\begin{thebibliography}{27}%
\makeatletter
\providecommand \@ifxundefined [1]{%
 \@ifx{#1\undefined}
}%
\providecommand \@ifnum [1]{%
 \ifnum #1\expandafter \@firstoftwo
 \else \expandafter \@secondoftwo
 \fi
}%
\providecommand \@ifx [1]{%
 \ifx #1\expandafter \@firstoftwo
 \else \expandafter \@secondoftwo
 \fi
}%
\providecommand \natexlab [1]{#1}%
\providecommand \enquote  [1]{``#1''}%
\providecommand \bibnamefont  [1]{#1}%
\providecommand \bibfnamefont [1]{#1}%
\providecommand \citenamefont [1]{#1}%
\providecommand \href@noop [0]{\@secondoftwo}%
\providecommand \href [0]{\begingroup \@sanitize@url \@href}%
\providecommand \@href[1]{\@@startlink{#1}\@@href}%
\providecommand \@@href[1]{\endgroup#1\@@endlink}%
\providecommand \@sanitize@url [0]{\catcode `\\12\catcode `\$12\catcode
  `\&12\catcode `\#12\catcode `\^12\catcode `\_12\catcode `\%12\relax}%
\providecommand \@@startlink[1]{}%
\providecommand \@@endlink[0]{}%
\providecommand \url  [0]{\begingroup\@sanitize@url \@url }%
\providecommand \@url [1]{\endgroup\@href {#1}{\urlprefix }}%
\providecommand \urlprefix  [0]{URL }%
\providecommand \Eprint [0]{\href }%
\providecommand \doibase [0]{http://dx.doi.org/}%
\providecommand \selectlanguage [0]{\@gobble}%
\providecommand \bibinfo  [0]{\@secondoftwo}%
\providecommand \bibfield  [0]{\@secondoftwo}%
\providecommand \translation [1]{[#1]}%
\providecommand \BibitemOpen [0]{}%
\providecommand \bibitemStop [0]{}%
\providecommand \bibitemNoStop [0]{.\EOS\space}%
\providecommand \EOS [0]{\spacefactor3000\relax}%
\providecommand \BibitemShut  [1]{\csname bibitem#1\endcsname}%
\let\auto@bib@innerbib\@empty
\bibitem [{\citenamefont {Fudenberg}\ and\ \citenamefont
  {Tirole}(1991)}]{fudenberg1991game}%
  \BibitemOpen
  \bibfield  {author} {\bibinfo {author} {\bibfnamefont {D.}~\bibnamefont
  {Fudenberg}}\ and\ \bibinfo {author} {\bibfnamefont {J.}~\bibnamefont
  {Tirole}},\ }\href@noop {} {\bibfield  {journal} {\bibinfo  {journal} {MIT
  Press}\ } (\bibinfo {year} {1991})}\BibitemShut {NoStop}%
\bibitem [{\citenamefont {Smith}\ and\ \citenamefont
  {Price}(1973)}]{smith1973logic}%
  \BibitemOpen
  \bibfield  {author} {\bibinfo {author} {\bibfnamefont {J.~M.}\ \bibnamefont
  {Smith}}\ and\ \bibinfo {author} {\bibfnamefont {G.~R.}\ \bibnamefont
  {Price}},\ }\href@noop {} {\bibfield  {journal} {\bibinfo  {journal}
  {Nature}\ }\textbf {\bibinfo {volume} {246}},\ \bibinfo {pages} {15}
  (\bibinfo {year} {1973})}\BibitemShut {NoStop}%
\bibitem [{\citenamefont {Smith}(1982)}]{smith1982evolution}%
  \BibitemOpen
  \bibfield  {author} {\bibinfo {author} {\bibfnamefont {J.~M.}\ \bibnamefont
  {Smith}},\ }\href@noop {} {\emph {\bibinfo {title} {Evolution and the Theory
  of Games}}}\ (\bibinfo  {publisher} {Cambridge university press},\ \bibinfo
  {year} {1982})\BibitemShut {NoStop}%
\bibitem [{\citenamefont {Nowak}\ and\ \citenamefont
  {Sigmund}(2004)}]{nowak2004evolutionary}%
  \BibitemOpen
  \bibfield  {author} {\bibinfo {author} {\bibfnamefont {M.~A.}\ \bibnamefont
  {Nowak}}\ and\ \bibinfo {author} {\bibfnamefont {K.}~\bibnamefont
  {Sigmund}},\ }\href@noop {} {\bibfield  {journal} {\bibinfo  {journal}
  {Science}\ }\textbf {\bibinfo {volume} {303}},\ \bibinfo {pages} {793}
  (\bibinfo {year} {2004})}\BibitemShut {NoStop}%
\bibitem [{\citenamefont {Kreps}(1990)}]{kreps1990game}%
  \BibitemOpen
  \bibfield  {author} {\bibinfo {author} {\bibfnamefont {D.~M.}\ \bibnamefont
  {Kreps}},\ }\href@noop {} {\emph {\bibinfo {title} {Game theory and economic
  modelling}}}\ (\bibinfo  {publisher} {Oxford University Press},\ \bibinfo
  {year} {1990})\BibitemShut {NoStop}%
\bibitem [{\citenamefont {Van~der Ploeg}\ and\ \citenamefont
  {de~Zeeuw}(2016)}]{van2016non}%
  \BibitemOpen
  \bibfield  {author} {\bibinfo {author} {\bibfnamefont {F.}~\bibnamefont
  {Van~der Ploeg}}\ and\ \bibinfo {author} {\bibfnamefont {A.}~\bibnamefont
  {de~Zeeuw}},\ }\href@noop {} {\bibfield  {journal} {\bibinfo  {journal}
  {Environmental and resource economics}\ }\textbf {\bibinfo {volume} {65}},\
  \bibinfo {pages} {519} (\bibinfo {year} {2016})}\BibitemShut {NoStop}%
\bibitem [{\citenamefont {Morrow}(1994)}]{morrow1994game}%
  \BibitemOpen
  \bibfield  {author} {\bibinfo {author} {\bibfnamefont {J.~D.}\ \bibnamefont
  {Morrow}},\ }\href@noop {} {\emph {\bibinfo {title} {Game theory for
  political scientists}}},\ \bibinfo {number} {30: 519.83}\ (\bibinfo
  {publisher} {Princeton University Press,},\ \bibinfo {year}
  {1994})\BibitemShut {NoStop}%
\bibitem [{\citenamefont {Buskens}\ and\ \citenamefont
  {Snijders}(2016)}]{buskens2016effects}%
  \BibitemOpen
  \bibfield  {author} {\bibinfo {author} {\bibfnamefont {V.}~\bibnamefont
  {Buskens}}\ and\ \bibinfo {author} {\bibfnamefont {C.}~\bibnamefont
  {Snijders}},\ }\href@noop {} {\bibfield  {journal} {\bibinfo  {journal}
  {Dynamic games and applications}\ }\textbf {\bibinfo {volume} {6}},\ \bibinfo
  {pages} {477} (\bibinfo {year} {2016})}\BibitemShut {NoStop}%
\bibitem [{\citenamefont {MacLean}\ and\ \citenamefont
  {Gudelj}(2006)}]{maclean2006resource}%
  \BibitemOpen
  \bibfield  {author} {\bibinfo {author} {\bibfnamefont {R.~C.}\ \bibnamefont
  {MacLean}}\ and\ \bibinfo {author} {\bibfnamefont {I.}~\bibnamefont
  {Gudelj}},\ }\href@noop {} {\bibfield  {journal} {\bibinfo  {journal}
  {Nature}\ }\textbf {\bibinfo {volume} {441}},\ \bibinfo {pages} {498}
  (\bibinfo {year} {2006})}\BibitemShut {NoStop}%
\bibitem [{\citenamefont {Rand}\ \emph {et~al.}(2011)\citenamefont {Rand},
  \citenamefont {Arbesman},\ and\ \citenamefont
  {Christakis}}]{rand2011dynamic}%
  \BibitemOpen
  \bibfield  {author} {\bibinfo {author} {\bibfnamefont {D.~G.}\ \bibnamefont
  {Rand}}, \bibinfo {author} {\bibfnamefont {S.}~\bibnamefont {Arbesman}}, \
  and\ \bibinfo {author} {\bibfnamefont {N.~A.}\ \bibnamefont {Christakis}},\
  }\href@noop {} {\bibfield  {journal} {\bibinfo  {journal} {Proceedings of the
  National Academy of Sciences}\ }\textbf {\bibinfo {volume} {108}},\ \bibinfo
  {pages} {19193} (\bibinfo {year} {2011})}\BibitemShut {NoStop}%
\bibitem [{\citenamefont {Gracia-L{\'a}zaro}\ \emph {et~al.}(2012)\citenamefont
  {Gracia-L{\'a}zaro}, \citenamefont {Ferrer}, \citenamefont {Ruiz},
  \citenamefont {Taranc{\'o}n}, \citenamefont {Cuesta}, \citenamefont
  {S{\'a}nchez},\ and\ \citenamefont {Moreno}}]{gracia2012heterogeneous}%
  \BibitemOpen
  \bibfield  {author} {\bibinfo {author} {\bibfnamefont {C.}~\bibnamefont
  {Gracia-L{\'a}zaro}}, \bibinfo {author} {\bibfnamefont {A.}~\bibnamefont
  {Ferrer}}, \bibinfo {author} {\bibfnamefont {G.}~\bibnamefont {Ruiz}},
  \bibinfo {author} {\bibfnamefont {A.}~\bibnamefont {Taranc{\'o}n}}, \bibinfo
  {author} {\bibfnamefont {J.~A.}\ \bibnamefont {Cuesta}}, \bibinfo {author}
  {\bibfnamefont {A.}~\bibnamefont {S{\'a}nchez}}, \ and\ \bibinfo {author}
  {\bibfnamefont {Y.}~\bibnamefont {Moreno}},\ }\href@noop {} {\bibfield
  {journal} {\bibinfo  {journal} {Proceedings of the National Academy of
  Sciences}\ }\textbf {\bibinfo {volume} {109}},\ \bibinfo {pages} {12922}
  (\bibinfo {year} {2012})}\BibitemShut {NoStop}%
\bibitem [{\citenamefont {Turner}\ and\ \citenamefont
  {Chao}(1999)}]{turner1999prisoner}%
  \BibitemOpen
  \bibfield  {author} {\bibinfo {author} {\bibfnamefont {P.~E.}\ \bibnamefont
  {Turner}}\ and\ \bibinfo {author} {\bibfnamefont {L.}~\bibnamefont {Chao}},\
  }\href@noop {} {\bibfield  {journal} {\bibinfo  {journal} {Nature}\ }\textbf
  {\bibinfo {volume} {398}},\ \bibinfo {pages} {441} (\bibinfo {year}
  {1999})}\BibitemShut {NoStop}%
\bibitem [{\citenamefont {Gore}\ \emph {et~al.}(2009)\citenamefont {Gore},
  \citenamefont {Youk},\ and\ \citenamefont
  {Van~Oudenaarden}}]{gore2009snowdrift}%
  \BibitemOpen
  \bibfield  {author} {\bibinfo {author} {\bibfnamefont {J.}~\bibnamefont
  {Gore}}, \bibinfo {author} {\bibfnamefont {H.}~\bibnamefont {Youk}}, \ and\
  \bibinfo {author} {\bibfnamefont {A.}~\bibnamefont {Van~Oudenaarden}},\
  }\href@noop {} {\bibfield  {journal} {\bibinfo  {journal} {Nature}\ }\textbf
  {\bibinfo {volume} {459}},\ \bibinfo {pages} {253} (\bibinfo {year}
  {2009})}\BibitemShut {NoStop}%
\bibitem [{\citenamefont {Zomorrodi}\ and\ \citenamefont
  {Segr{\`e}}(2017)}]{zomorrodi2017genome}%
  \BibitemOpen
  \bibfield  {author} {\bibinfo {author} {\bibfnamefont {A.~R.}\ \bibnamefont
  {Zomorrodi}}\ and\ \bibinfo {author} {\bibfnamefont {D.}~\bibnamefont
  {Segr{\`e}}},\ }\href@noop {} {\bibfield  {journal} {\bibinfo  {journal}
  {Nature Communications}\ }\textbf {\bibinfo {volume} {8}},\ \bibinfo {pages}
  {1563} (\bibinfo {year} {2017})}\BibitemShut {NoStop}%
\bibitem [{\citenamefont {Oliveira}\ \emph {et~al.}(2014)\citenamefont
  {Oliveira}, \citenamefont {Niehus},\ and\ \citenamefont
  {Foster}}]{oliveira2014evolutionary}%
  \BibitemOpen
  \bibfield  {author} {\bibinfo {author} {\bibfnamefont {N.~M.}\ \bibnamefont
  {Oliveira}}, \bibinfo {author} {\bibfnamefont {R.}~\bibnamefont {Niehus}}, \
  and\ \bibinfo {author} {\bibfnamefont {K.~R.}\ \bibnamefont {Foster}},\
  }\href@noop {} {\bibfield  {journal} {\bibinfo  {journal} {Proceedings of the
  National Academy of Sciences}\ }\textbf {\bibinfo {volume} {111}},\ \bibinfo
  {pages} {17941} (\bibinfo {year} {2014})}\BibitemShut {NoStop}%
\bibitem [{\citenamefont {Pollock}(1994)}]{pollock1994social}%
  \BibitemOpen
  \bibfield  {author} {\bibinfo {author} {\bibfnamefont {G.~B.}\ \bibnamefont
  {Pollock}},\ }\href@noop {} {\bibfield  {journal} {\bibinfo  {journal}
  {Evolutionary ecology}\ }\textbf {\bibinfo {volume} {8}},\ \bibinfo {pages}
  {221} (\bibinfo {year} {1994})}\BibitemShut {NoStop}%
\bibitem [{\citenamefont {Nash}\ \emph {et~al.}(1950)\citenamefont {Nash} \emph
  {et~al.}}]{nash1950equilibrium}%
  \BibitemOpen
  \bibfield  {author} {\bibinfo {author} {\bibfnamefont {J.~F.}\ \bibnamefont
  {Nash}} \emph {et~al.},\ }\href@noop {} {\bibfield  {journal} {\bibinfo
  {journal} {Proceedings of the national academy of sciences}\ }\textbf
  {\bibinfo {volume} {36}},\ \bibinfo {pages} {48} (\bibinfo {year}
  {1950})}\BibitemShut {NoStop}%
\bibitem [{\citenamefont {Aumann}(1987)}]{aumann1987correlated}%
  \BibitemOpen
  \bibfield  {author} {\bibinfo {author} {\bibfnamefont {R.~J.}\ \bibnamefont
  {Aumann}},\ }\href@noop {} {\bibfield  {journal} {\bibinfo  {journal}
  {Econometrica: Journal of the Econometric Society}\ ,\ \bibinfo {pages} {1}}
  (\bibinfo {year} {1987})}\BibitemShut {NoStop}%
\bibitem [{\citenamefont {Stoof}\ \emph {et~al.}(2009)\citenamefont {Stoof},
  \citenamefont {Gubbels},\ and\ \citenamefont
  {Dickerscheid}}]{stoof2009ultracold}%
  \BibitemOpen
  \bibfield  {author} {\bibinfo {author} {\bibfnamefont {H.~T.}\ \bibnamefont
  {Stoof}}, \bibinfo {author} {\bibfnamefont {K.~B.}\ \bibnamefont {Gubbels}},
  \ and\ \bibinfo {author} {\bibfnamefont {D.}~\bibnamefont {Dickerscheid}},\
  }\href@noop {} {\emph {\bibinfo {title} {Ultracold quantum fields}}}\
  (\bibinfo  {publisher} {Springer},\ \bibinfo {year} {2009})\BibitemShut
  {NoStop}%
\bibitem [{\citenamefont {Mailath}\ \emph {et~al.}(1997)\citenamefont
  {Mailath}, \citenamefont {Samuelson},\ and\ \citenamefont
  {Shaked}}]{mailath1997correlated}%
  \BibitemOpen
  \bibfield  {author} {\bibinfo {author} {\bibfnamefont {G.~J.}\ \bibnamefont
  {Mailath}}, \bibinfo {author} {\bibfnamefont {L.}~\bibnamefont {Samuelson}},
  \ and\ \bibinfo {author} {\bibfnamefont {A.}~\bibnamefont {Shaked}},\
  }\href@noop {} {\bibfield  {journal} {\bibinfo  {journal} {Economic Theory}\
  }\textbf {\bibinfo {volume} {9}},\ \bibinfo {pages} {551} (\bibinfo {year}
  {1997})}\BibitemShut {NoStop}%
\bibitem [{\citenamefont {Wong}\ \emph {et~al.}(2004)\citenamefont {Wong},
  \citenamefont {Kim} \emph {et~al.}}]{wong2004evolutionarily}%
  \BibitemOpen
  \bibfield  {author} {\bibinfo {author} {\bibfnamefont {K.-C.}\ \bibnamefont
  {Wong}}, \bibinfo {author} {\bibfnamefont {C.}~\bibnamefont {Kim}},  \emph
  {et~al.},\ }in\ \href@noop {} {\emph {\bibinfo {booktitle} {Econometric
  Society 2004 Far Eastern Meetings}}},\ \bibinfo {series and number} {\bibinfo
  {number} {495}}\ (\bibinfo {organization} {Econometric Society},\ \bibinfo
  {year} {2004})\BibitemShut {NoStop}%
\bibitem [{\citenamefont {Lenzo}\ and\ \citenamefont
  {Sarver}(2006)}]{lenzo2006correlated}%
  \BibitemOpen
  \bibfield  {author} {\bibinfo {author} {\bibfnamefont {J.}~\bibnamefont
  {Lenzo}}\ and\ \bibinfo {author} {\bibfnamefont {T.}~\bibnamefont {Sarver}},\
  }\href@noop {} {\bibfield  {journal} {\bibinfo  {journal} {Games and Economic
  Behavior}\ }\textbf {\bibinfo {volume} {56}},\ \bibinfo {pages} {271}
  (\bibinfo {year} {2006})}\BibitemShut {NoStop}%
\bibitem [{\citenamefont {Metzger}(2018)}]{metzger2018evolution}%
  \BibitemOpen
  \bibfield  {author} {\bibinfo {author} {\bibfnamefont {L.~P.}\ \bibnamefont
  {Metzger}},\ }\href@noop {} {\bibfield  {journal} {\bibinfo  {journal}
  {Journal of Evolutionary Economics}\ }\textbf {\bibinfo {volume} {28}},\
  \bibinfo {pages} {333} (\bibinfo {year} {2018})}\BibitemShut {NoStop}%
\bibitem [{\citenamefont {Cripps}(1991)}]{cripps1991correlated}%
  \BibitemOpen
  \bibfield  {author} {\bibinfo {author} {\bibfnamefont {M.}~\bibnamefont
  {Cripps}},\ }\href@noop {} {\bibfield  {journal} {\bibinfo  {journal}
  {Journal of Economic Theory}\ }\textbf {\bibinfo {volume} {55}},\ \bibinfo
  {pages} {428} (\bibinfo {year} {1991})}\BibitemShut {NoStop}%
\bibitem [{\citenamefont {Cassar}(2007)}]{cassar2007coordination}%
  \BibitemOpen
  \bibfield  {author} {\bibinfo {author} {\bibfnamefont {A.}~\bibnamefont
  {Cassar}},\ }\href@noop {} {\bibfield  {journal} {\bibinfo  {journal} {Games
  and Economic Behavior}\ }\textbf {\bibinfo {volume} {58}},\ \bibinfo {pages}
  {209} (\bibinfo {year} {2007})}\BibitemShut {NoStop}%
\bibitem [{\citenamefont {Broere}\ \emph {et~al.}(2017)\citenamefont {Broere},
  \citenamefont {Buskens}, \citenamefont {Weesie},\ and\ \citenamefont
  {Stoof}}]{broere2017network}%
  \BibitemOpen
  \bibfield  {author} {\bibinfo {author} {\bibfnamefont {J.}~\bibnamefont
  {Broere}}, \bibinfo {author} {\bibfnamefont {V.}~\bibnamefont {Buskens}},
  \bibinfo {author} {\bibfnamefont {J.}~\bibnamefont {Weesie}}, \ and\ \bibinfo
  {author} {\bibfnamefont {H.}~\bibnamefont {Stoof}},\ }\href@noop {}
  {\bibfield  {journal} {\bibinfo  {journal} {Scientific reports}\ }\textbf
  {\bibinfo {volume} {7}},\ \bibinfo {pages} {17016} (\bibinfo {year}
  {2017})}\BibitemShut {NoStop}%
\bibitem [{\citenamefont {Adami}\ and\ \citenamefont
  {Hintze}(2018)}]{adami2018thermodynamics}%
  \BibitemOpen
  \bibfield  {author} {\bibinfo {author} {\bibfnamefont {C.}~\bibnamefont
  {Adami}}\ and\ \bibinfo {author} {\bibfnamefont {A.}~\bibnamefont {Hintze}},\
  }\href@noop {} {\bibfield  {journal} {\bibinfo  {journal} {Physical Review
  E}\ }\textbf {\bibinfo {volume} {97}},\ \bibinfo {pages} {062136} (\bibinfo
  {year} {2018})}\BibitemShut {NoStop}%
\end{thebibliography}%

\end{document}